\documentclass[preprint,12pt]{elsarticle}

\usepackage{amssymb}
\usepackage{amsthm}
\usepackage{amsmath,amsfonts}
\usepackage{multirow}
\usepackage{tabularx}
\usepackage{caption}
\usepackage{subcaption}
\usepackage{float}
\usepackage{booktabs}
\usepackage[utf8]{inputenc}
\usepackage[T1]{fontenc}
\usepackage{lmodern}
\usepackage{xcolor}
\usepackage{soul}
\usepackage{threeparttable}
\usepackage{textcomp}
\usepackage{mathrsfs}
\usepackage[english]{babel}
\usepackage[colorlinks,citecolor=red,urlcolor=blue,bookmarks=false,hypertexnames=true]{hyperref}
\usepackage[export]{adjustbox}

\newcommand{\kJ}{\,kJ\,mol$^{-1}$}
\newcommand{\Xb}{\bar X}
\newcommand{\gb}{\mathrm{gb}}

\journal{Acta Materialia}

\begin{document}
\emergencystretch 3em

\begin{frontmatter}

\title{A macroscopic-atom spectral theory of grain-boundary segregation in multicomponent alloys}

\author[a]{Dennis Boakye}
\author[a]{Chuang Deng \corref{b}}
\cortext[b]{Corresponding author}
\ead{chuang.deng@umanitoba.ca}

\affiliation[a]{organization={Mechanical Engineering, University of Manitoba},
            addressline={66 Chancellors Cir},
            city={Winnipeg},
            postcode={R3T 2N2},
            state={Manitoba},
            country={Canada}}

\begin{abstract}
Predicting grain-boundary segregation in multicomponent alloys remains challenging because existing spectral approaches require alloy-specific atomistic calculations to resolve segregation-energy distributions and correlations across non-equivalent boundary sites. Here, we develop a transferable macroscopic-atom spectral theory that predicts the joint segregation behavior of multiple species from physically interpretable elemental descriptors. The theory is founded on the finding that local strain and bond deficit account for 91\% of the site-to-site co-variation of 40 solutes in a density-functional-theory (DFT) Al polycrystal. Linked to atomic size and surface enthalpy, these coordinates allow prediction of individual segregation spectra and their cross-species correlations, which determine whether species compete for or share grain-boundary sites. A closed-system formulation unifies host-element redistribution and dilute-solute segregation, couples metallic and light-element sublattices, and accounts for bulk compound formation and solute pairing. Calibrated using binary spectra only, the theory predicts unseen-host spectral means with a correlation coefficient of 0.81, outperforms embedded-atom spectra against DFT benchmarks, and ranks site competition across 870 solute pairs with a Spearman coefficient of 0.90. Without fitting to multicomponent observations, it reproduces 11 of 12 equilibrium Fe ternaries and 49 of 55 Mg ternaries. Predictions recorded before simulations of an 82,699-atom polycrystal identify site competition as the baseline behavior and metal-metalloid pair chemistry as a source of deviations requiring atomistic resolution. These results establish a physically grounded route to screening grain-boundary chemistry in multicomponent alloys, including steels and high-entropy alloys.
\end{abstract}


\end{frontmatter}


\section{Introduction}
\label{sec:intro}

Grain boundaries hold a small fraction of the atoms in a polycrystal, yet they often decide whether an alloy is tough or brittle, whether it corrodes, and whether a fine microstructure survives heating. They do so largely through chemistry. A few hundred parts per million of phosphorus or tin, collected at the boundaries of a steel during tempering or slow cooling, can turn a ductile fracture into an intergranular one, while an addition that segregates to the same boundaries can lower their energy and keep a nanocrystalline alloy stable at high temperature~\cite{lejcek2010grain,lejcek2017interfacial,raabe2014grain,weissmuller1993alloy,kirchheim2002grain,chookajorn2012design}. Which of these happens depends on which atoms the boundary takes up, and in a real alloy that is the outcome of a contest. Every species, including the principal elements, competes for a limited set of boundary sites that differ from one another, and each species changes what the others gain by being there. For an alloy that has not yet been made, predicting the outcome of this contest is a central challenge of interface thermodynamics, and it is the problem we seek to solve.

The classical theory of segregation provides the thermodynamic framework. McLean's isotherm describes a solute exchanging between the bulk and a single kind of boundary site~\cite{mclean1957grain}. Fowler and Guggenheim added interactions between the adsorbed atoms~\cite{fowler1939statistical}, Hondros and Seah drew the analogy with surface adsorption~\cite{hondros1977theory}, and Guttmann showed how the affinity between an impurity and an alloying element raises or lowers the segregation of the impurity in a ternary alloy, which explained temper embrittlement~\cite{guttmann1975equilibrium,guttmann1995thermochemical}. Wynblatt and Ku separated the segregation energy into a surface term, an elastic misfit term and a chemical term~\cite{wynblatt1977surface}. These descriptions, though identify the right physics, treat the boundary as one kind of site whereas a real polycrystalline boundary is not.

In the spectral description, the single energy is replaced by a distribution of segregation energies over the non-equivalent boundary sites~\cite{white1977spectrum,kirchheim1988hydrogen,huber2017abinitio,wagih2019spectrum,hu2024computational,matson2026overview}. Wagih et al. \cite{wagih2020learning} computed such spectra for hundreds of binary alloys, brought them to density-functional (DFT) accuracy~\cite{wagih2022learning}, and extended them to finite concentrations~\cite{wagih2020grain}, interstitial solutes~\cite{wagih2023spectrum} and the stability of nanocrystalline alloys~\cite{wagih2021thermodynamics}. In their recent work, the authors showed that when two solutes are present, the correlation between their spectra decides whether they either compete for the same sites or occupy different ones~\cite{wagih2025designing}. Spectral isotherm for dilute solutes without interactions has also been derived by Marchiy et al.~\cite{marchiy2025spectral,marchiy2026multicomponent}. Furthermore, Zhang and Deng mapped co-segregation in all 55 ternaries of 11 solutes in Mg from dual-solute spectra (DS) and hybrid molecular dynamics and Monte Carlo (MD/MC)~\cite{zhang2026predicting}. Moreover, current frameworks using descriptor models have linked segregation energies to atomic size, coordination and electronic structure~\cite{wu2025unconventional,li2024analytic,li2025universal,li2026alloying}.

However, in each of these approaches, the spectra of an alloy are computed for that alloy. Each requires an interatomic potential or a DFT calculation for the particular combination of elements, and the computational cost grows quickly with the number of components. For instance, an alloy with $n$ species has $n$ spectra and $n(n-1)/2$ pairwise correlations, besides the chemical interactions between the species. Moreover, reliable potentials do not exist for most combinations of transition metals (TM), metalloids and light elements. In a high-entropy alloy (HEA) the host itself is a mixture whose elements redistribute at the boundary, and in a steel the impurity that segregates is only the part that has not been tied up in sulfides, nitrides, phosphides or carbides inside the grains. A theory that could predict the joint spectrum of an arbitrary alloy, and solve the resulting competition with these effects included, is still missing.

The macroscopic-atom model (MAM) of Miedema and co-workers provides the physical basis for such a theory~\cite{miedema1980cohesion,deboer1988cohesion}. The MAM describes a metal as a packing of atomic cells, each carrying the electron density ($n_{ws}$) and electronegativity ($\phi*$) of the pure element, and expressed the energy of an alloy as the energy of the contacts between dissimilar cells. Three quantities fix that energy: a charge-transfer attraction set by the difference in $\phi^*$, a repulsion from the mismatch of $n_{ws}$ where two cells meet, and the surface enthalpy of the cells, all scaled by the contact area $V^{2/3}$. A grain boundary is a region in which exactly these contacts are disturbed. An atom on a boundary site has lost part of its contact area, sits in a cell that is stretched or compressed, and meets different neighbors. The energy it gains or loses there is therefore controlled by its surface enthalpy and size relative to the host, its chemical interaction with the host and with the other segregated atoms. These are the surface, elastic and chemical contributions of Wynblatt and Ku and the interaction parameters of Guttmann, now available from one consistent set of tabulated parameters for almost every metal and, through contact fractions, for concentrated alloys. Miedema-type estimates of the segregation enthalpy have been used to screen thousands of binary alloys for nanocrystalline stability~\cite{murdoch2013estimation}, and in our recent paper, we employed the same framework to predict segregation and adhesion at metal-oxide interfaces in HEAs~\cite{boakye2027rapid}. However, the macroscopic-atom picture lacks the variation from one boundary site to the next and the correlation of that variation between species. Our central observation is that this variation is determined by the same quantities. A boundary site that is dilated is dilated for every atom placed on it, and a site that is short of neighbors is short of them for every atom. How strongly a species responds to such a site depends on its size and its surface enthalpy, the quantities that also set its mean segregation energy. The macroscopic-atom descriptors can therefore predict not only where the spectrum of each species lies, but also how wide it is and how it correlates with the spectra of all the others.

In this work, we turn that observation into a predictive theory of multicomponent grain-boundary segregation, which is summarized in Fig.~\ref{fig:overview}. The theory predicts the complete joint spectrum of an alloy from tabulated elemental properties, without an atomistic calculation of that alloy. It reduces the site-to-site variation of all species to two physical coordinates, local strain and bond deficit, which explains why some species compete for the same sites while others share the boundary. It references host elements and dilute solutes to the same average atom, so that the redistribution of the principal elements of an HEA and the segregation of a trace addition are described by one isotherm. Furthermore, it couples the metallic boundary to a sublattice of light elements and to the bulk compounds and solute pairs that decide how much of each impurity is free to segregate. Finally, by comparing the theory with atomistic MC simulations on a realistic polycrystal, with the predictions recorded before the simulations, it separates the behavior that transferable thermodynamics can capture from the pair chemistry that requires atomistic resolution. The theory is calibrated on binary spectra alone and is tested against DFT data, exact site-competition calculations and multicomponent observations from experiments and simulations that played no part in its construction.

\section{Theory and methods}
\label{sec:theory}

\subsection{The boundary as a closed ensemble of sites}
\label{sec:iso}

We consider a polycrystal of mean grain size $d$. Energies are given per mole of atoms, $R$ is the gas constant and $T$ the absolute temperature, and all symbols are also listed in the Supplementary Material. The fraction of the atoms that belong to the boundaries is approximated by
\begin{equation}
f=1-\left(1-\frac{t}{d}\right)^3,
\label{eq:f}
\end{equation}
where $t=0.5$\,nm is an effective boundary thickness. Equation~\eqref{eq:f} is the only place where the grain size enters. It fixes how many boundary sites are available for each solute atom and turns a property of a boundary into a property of a polycrystal.

The boundary is depicted as a set of sites, labeled $i$, that differ in their local atomic environment. Every site holds exactly one atom. The species that can occupy it are labeled $a$ and $b$. Its worth noting that in a concentrated alloy they include the principal elements as well as the dilute additions, and both are treated in the same way. The segregation energy $E_{ia}$ is the energy change when an atom of $a$ moves from a site in the grain interior (bulk) to boundary site $i$ and the atom it replaces at $i$ moves into the bulk. It is measured relative to the average host atom $M$ defined in Section~\ref{sec:mean}, such that $E_{iM}=0$ for every site. A negative $E_{ia}$ means that site $i$ is more favorable for $a$ than a site inside a grain.

At equilibrium, the probability that site $i$ is occupied by species $a$ is
\begin{equation}
X_{ia}=\frac{X^c_a\exp\!\left[-\left(E_{ia}+\phi^{\gb}_a-\phi^c_a\right)/RT\right]}{\sum_bX^c_b\exp\!\left[-\left(E_{ib}+\phi^{\gb}_b-\phi^c_b\right)/RT\right]}.
\label{eq:iso}
\end{equation}
Here $X^c_a$ is the atom fraction of $a$ in the crystalline bulk. The grains are the reservoir with which the boundary exchanges atoms, and $X^c_a$ enters as the ideal activity of $a$ in that reservoir. The more of $a$ there is in the grains, the more often a boundary site is offered an atom of $a$. The quantities $\phi^c_a$ and $\phi^{\gb}_a$ are the excess (non-ideal) parts of the chemical potential of $a$ in the grains and at the boundary, and their difference is the free energy that an atom of $a$ gains or loses at the boundary because of the atoms already segregated there. The exponential is a Boltzmann factor that weights species $a$ by its site energy and by this interaction term. The sum in the denominator runs over all species and makes the occupancies of each site add up to one. It is also where site competition comes from. That is a species that binds strongly at site $i$ makes the denominator large and lowers the occupancy of every other species at that site, whether or not the two species interact.

In the regular-solution approximation the two excess chemical potentials are
\begin{equation}
\phi^c_a=\sum_b\Omega_{ab}\,X^c_b,\qquad \phi^{\gb}_a=\sum_b\left(\lambda\,\Omega_{ab}+W_{ab}\right)\Xb_b,
\label{eq:phi}
\end{equation}
where $\Xb_b$ is the boundary concentration of $b$, that is, $X_{ib}$ averaged over all boundary sites, $\Omega_{ab}$ is the regular-solution interaction parameter between $a$ and $b$, $\lambda$ is the ratio of the interaction at the boundary to that in the bulk, and $W_{ab}$ is an elastic interaction that acts only at the boundary (Section~\ref{sec:inter}). Negative values mean attraction analogous to the Warren-Cowley parameter for ordering. If $a$ and $b$ attract each other and $b$ is enriched at the boundary, $\phi^{\gb}_a$ is more negative than $\phi^c_a$ and $a$ is drawn to the boundary as well. The opposite happens with repulsion. Any contribution that is the same for all species cancels between numerator and denominator of Eq.~\eqref{eq:iso}, so only differences between species matter.

The polycrystal is closed, so atoms that move to the boundary leave the grains:
\begin{equation}
f\,\Xb_a+(1-f)\,X^c_a=X^{\rm tot}_a,
\label{eq:mass}
\end{equation}
where $X^{\rm tot}_a$ is the amount of $a$ per atom of alloy that is free to segregate. Without bulk sinks it is the overall composition. Equation~\eqref{eq:mass} closes the problem and introduces a second, indirect way in which species compete. That is a species that segregates strongly empties the grains and so changes the reservoir from which every other species is drawn, even if the two prefer different sites.

Equations~\eqref{eq:iso} to \eqref{eq:mass} are the conditions for a stationary point of the Gibbs energy per atom of the polycrystal at fixed overall composition,
\begin{multline}
G=(1-f)\Big[RT\sum_aX^c_a\ln X^c_a+\tfrac12\sum_{a,b}\Omega_{ab}X^c_aX^c_b\Big]\\
+f\Big[\Big\langle\sum_aE_{ia}X_{ia}+RT\sum_aX_{ia}\ln X_{ia}\Big\rangle_i+\tfrac12\sum_{a,b}\left(\lambda\Omega_{ab}+W_{ab}\right)\Xb_a\Xb_b\Big],
\label{eq:G}
\end{multline}
where $\langle\cdot\rangle_i$ denotes an average over the boundary sites. The first bracket is the free energy of the bulk matrix, made of the ideal entropy of mixing and the regular-solution enthalpy. The second is the free energy of the boundary comprising of the site energies, the configurational entropy of each site, and the interactions among the segregated atoms. The derivation is given in Supplementary Note~S1. With one solute, a single site energy and no interactions, Eq.~\eqref{eq:iso} reduces to the Langmuir-McLean isotherm. Adding interactions gives the Fowler-Guggenheim isotherm, one solute with a spectrum gives the White-Coghlan isotherm~\cite{white1977spectrum}, two interacting solutes with single site energies give Guttmann's ternary model~\cite{guttmann1975equilibrium}, and many solutes with spectra but no interactions give the dilute multicomponent isotherm of Marchiy et al.~\cite{marchiy2026multicomponent}.

Here, two quantities are used to report the results. The enrichment of species $a$ is
\begin{equation}
\beta_a=\frac{\Xb_a}{X^{\rm tot}_a},
\label{eq:beta}
\end{equation}
and a species is called segregated when $\beta_a>2$, depleted when $\log_{10}\beta_a<-0.1$, and neutral otherwise. The effect of adding species $b$ on the boundary concentration of species $a$ is measured by the coupling
\begin{equation}
C_{ab}=\ln\frac{\Xb_a\ (\text{with }b)}{\Xb_a\ (\text{without }b)}.
\label{eq:C}
\end{equation}
A positive $C_{ab}$ means co-segregation and a negative one antagonism. The coupling is not symmetric in general, because the influence of $b$ on $a$ need not equal $a$ on $b$. To understand the role of site preferences from that of chemistry, the equilibrium is solved again with $\Omega=W=0$. This gives the site part $C^{\rm site}_{ab}$, which reflects only the overlap of the spectra and the shared reservoir, and the difference $C^{\rm chem}_{ab}=C_{ab}-C^{\rm site}_{ab}$ is the chemical part. The split is a property of the model and not a unique experimental separation of mechanisms.

\begin{figure}[!ht]
	\centering
	\includegraphics[width=1\linewidth]{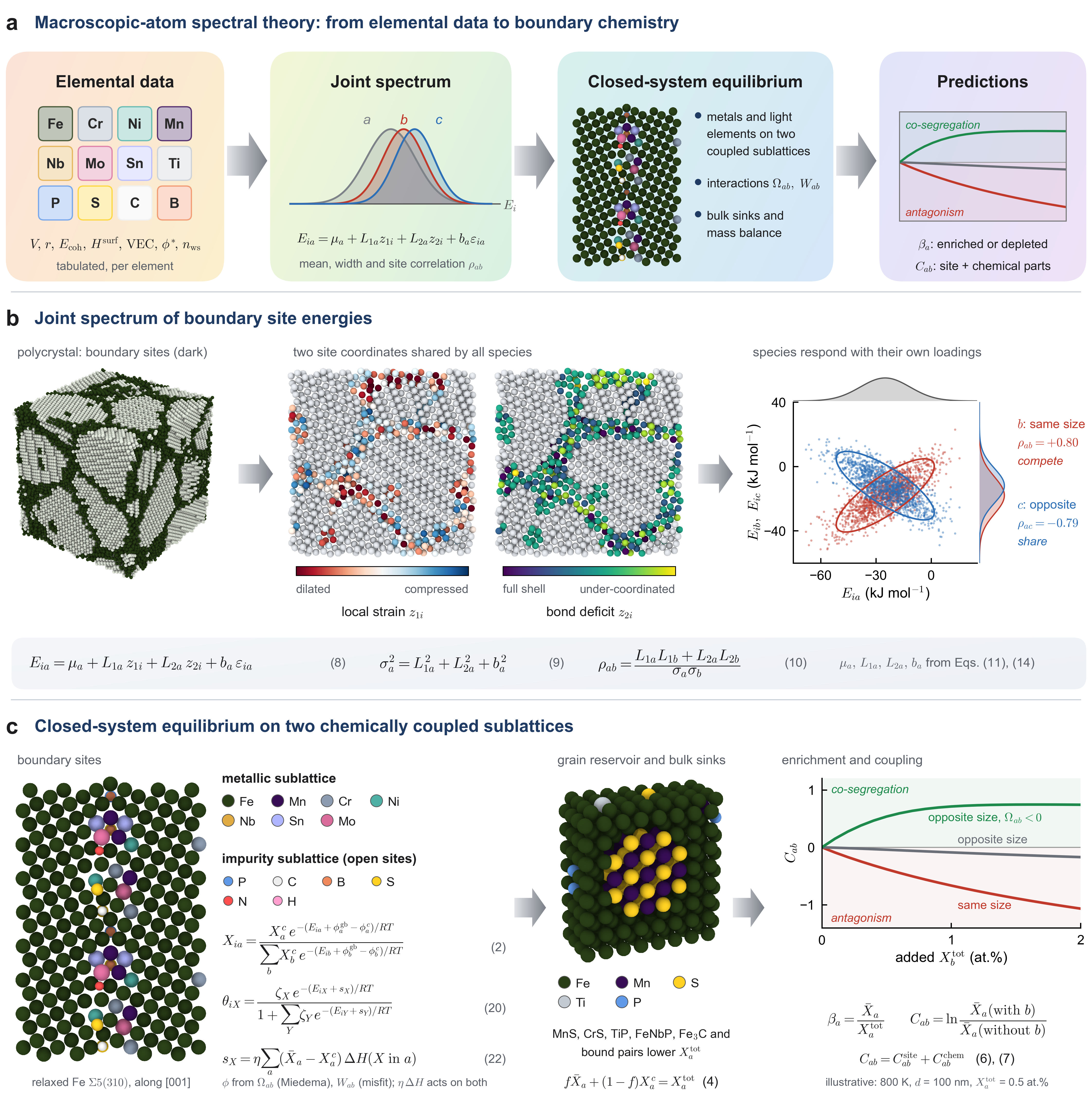}
	\caption{The macroscopic-atom spectral theory. (a) Workflow from tabulated elemental properties to the enrichment $\beta_a$ of every species and the coupling $C_{ab}$ between species. (b) The joint spectrum: boundary atoms of the 82,699-atom Fe polycrystal (dark green), the two shared site coordinates, local strain $z_{1i}$ and bond deficit $z_{2i}$, and illustrative site energies from Eq.~\eqref{eq:spec} for a partner $b$ that competes with $a$ ($\rho_{ab}=+0.80$) and a partner $c$ that shares the boundary with $a$ ($\rho_{ac}=-0.79$). (c) Closed-system equilibrium: metals and light elements on the two coupled sublattices of an Fe $\Sigma5(310)$ boundary, bulk sinks such as MnS that lower $X^{\rm tot}_a$, and the illustrative coupling $C_{ab}$ for partners of the same size, opposite size, and opposite size with $\Omega_{ab}<0$ (800\,K, 100\,nm grains).}
	\label{fig:overview}
\end{figure}

\subsection{The joint spectrum of boundary site energies}
\label{sec:joint}

We write the segregation energy of species $a$ at site $i$ as
\begin{equation}
E_{ia}=\mu_a+L_{1a}\,z_{1i}+L_{2a}\,z_{2i}+b_a\,\varepsilon_{ia},\qquad z_{1i},\,z_{2i},\,\varepsilon_{ia}\sim\mathcal N(0,1).
\label{eq:spec}
\end{equation}
The first term, $\mu_a$, is the mean segregation energy of $a$ over all sites. The three random variables describe how one site differs from another; each has zero mean and unit variance. The variables $z_{1i}$ and $z_{2i}$ belong to the site and take the same value for every species placed on it, whereas $\varepsilon_{ia}$ belongs to the combination of site and species.

The first shared coordinate, $z_{1i}$, is the local strain at site $i$, running from dilated to compressed environments. Its sign is chosen so that $z_{1i}>0$ at compressed sites and $z_{1i}<0$ at dilated ones. An oversized atom is stabilized at dilated sites and destabilized at compressed ones, and an undersized atom does the reverse. The loading $L_{1a}$, in kJ\,mol$^{-1}$, sets how strongly and in which direction species $a$ responds; it is positive for oversized and negative for undersized species. The second shared coordinate, $z_{2i}$, is the bond deficit of site $i$, that is, how far its coordination falls short of the crystal, with $z_{2i}>0$ at strongly under-coordinated sites. Species that lose little bonding energy when neighbors are missing respond to it differently from species that need a full coordination shell, and the loading $L_{2a}$ carries this difference. The last term, $b_a\varepsilon_{ia}$, collects the site-to-site variation that is specific to $a$, for example from its local chemical environment.

Because the random variables are independent and standardized, the spectrum of each species is a Gaussian with mean $\mu_a$ and variance
\begin{equation}
\sigma_a^2=L_{1a}^2+L_{2a}^2+b_a^2.
\label{eq:sigma}
\end{equation}
The shared coordinates produce a covariance between the energies of two species at the same site, and their site-by-site correlation is
\begin{equation}
\rho_{ab}=\frac{L_{1a}L_{1b}+L_{2a}L_{2b}}{\sigma_a\,\sigma_b}.
\label{eq:rho}
\end{equation}
A positive $\rho_{ab}$ means that the two species tend to find the same sites favorable and therefore compete for them; a typical case is two oversized solutes. A negative $\rho_{ab}$ means that the sites favorable for one are unfavorable for the other, so that the two can occupy the boundary side by side; an oversized and an undersized solute are the typical case (Fig.~\ref{fig:overview}b). In all, Eq.~\eqref{eq:spec} defines a multivariate normal distribution of the site energies of all species with means $\mu_a$ and covariance matrix $\Sigma_{ab}=L_{1a}L_{1b}+L_{2a}L_{2b}+\delta_{ab}b_a^2$, where $\delta_{ab}$ is the Kronecker delta. We call it the joint spectrum. We do not claim that a boundary site is fully characterized by two numbers, only that most of the correlation between species is and Section~\ref{sec:res_joint} tests this directly. Because the correlations depend on the loadings only through the inner products in Eq.~\eqref{eq:rho}, the two coordinates are defined up to a rotation. We name them after the descriptors that set their loadings (Section~\ref{sec:width}); Section~\ref{sec:res_joint} shows that the plane they span is the plane of local dilation and bond deficit, while each coordinate taken alone mixes the two.

\subsection{Mean segregation energy from macroscopic-atom descriptors}
\label{sec:mean}

The host is represented by a composition-weighted average atom $M$. Its molar volume $V_M$, metallic radius $r_M$, cohesive energy $E_{{\rm coh},M}$ and valence electron count are composition-weighted averages over the host elements, and its surface enthalpy is averaged with the Miedema contact fractions $s_k=X_kV_k^{2/3}/\sum_lX_lV_l^{2/3}$~\cite{deboer1988cohesion,boakye2027rapid}. A principal element of a concentrated alloy is then treated as a species relative to $M$, exactly as a dilute solute is. The mean segregation energy is a linear combination of six descriptors,
\begin{equation}
\mu_a=c_s\,\Delta H^{\rm surf}_a+c_e\,E_{{\rm coh},M}\,\delta_a^2+c_h\,\Delta H^\circ_{a\,{\rm in}\,M}+c_w\,w_a+c_{|w|}\,|w_a|+c_v\left[g(N_{d,a})-\bar g_M\right],
\label{eq:mu}
\end{equation}
whose coefficients are fitted once to binary spectra (Section~\ref{sec:calib}).

The surface descriptor $\Delta H^{\rm surf}_a$ is the Miedema surface enthalpy of $a$ minus the contact-fraction-weighted surface enthalpy of $M$. The surface enthalpy of an element, $H^{\rm surf}=C_0\gamma V^{2/3}$, is the energy needed to create the surface occupied by one mole of its atoms, where $\gamma$ is the surface energy of the solid, $V^{2/3}$ the molar surface area and $C_0$ a constant of the Miedema model~\cite{deboer1988cohesion}. It measures how much an atom loses when some of its bonds are broken. A boundary contains broken and weakened bonds, so an element with a lower surface enthalpy than the host loses less energy at a boundary site than the host atom it replaces, and its segregation lowers the energy of the polycrystal. This is the broken-bond contribution of the classical models~\cite{wynblatt1977surface,hondros1977theory}; the fitted $c_s$ is positive, as required by the argument.

The second descriptor is the elastic misfit energy. With the relative volume mismatch $\delta_a=(V_a-V_M)/V_M$, the product $E_{{\rm coh},M}\,\delta_a^2$ estimates the strain energy of an atom of the wrong size in the host lattice, with the cohesive energy of the host standing in for its stiffness. A boundary accommodates part of this strain, so a misfitting atom gains energy by moving there and the fitted $c_e$ is negative. Because $\delta_a$ is squared, the term depends only on the magnitude of the misfit.

The third descriptor, $\Delta H^\circ_{a\,{\rm in}\,M}$, is the Miedema enthalpy of solution of $a$ at infinite dilution in the host, averaged over the host elements with their contact fractions. Its fitted coefficient is close to zero for the binary calibration set. Chemistry enters the multicomponent problem mainly through the pair interactions of Section~\ref{sec:inter} rather than through the mean energy of a single solute.

The fourth and fifth descriptors carry the sign of the size difference. Following the strained-coordination analysis of Wu et al.~\cite{wu2025unconventional} we adopt
\begin{equation}
w_a=\frac{1}{12}\,E_{{\rm coh},M}\,\left(30-{\rm CN}_M\right)\left(\frac{r_a}{r_M}-1\right),
\label{eq:w}
\end{equation}
where $r_a$ and $r_M$ are the metallic radii of $a$ and of the average host atom and ${\rm CN}_M$ is the coordination number of the host lattice (12 for fcc and hcp, 8 for bcc). In their analysis of polycrystals of 16 elements, the segregation energy of a solute varies linearly with a strained coordination number of the site, and $w_a$ is the slope of that variation: it measures how strongly a solute of radius $r_a$ responds when the bonds around it are stretched or compressed. Its sign separates oversized ($w_a>0$) from undersized ($w_a<0$) solutes, and its magnitude grows with the stiffness of the host. The two terms $c_w\,w_a$ and $c_{|w|}\,|w_a|$ allow the response to be asymmetric. With the fitted coefficients both signs lower the mean energy, and an oversized solute gains about twice as much as an undersized one with the same $|w_a|$, consistent with the positive excess volume of grain boundaries, which makes dilated sites more common than compressed ones.

The sixth descriptor represents the $d$-band contribution to bonding. In Friedel's rectangular-band model~\cite{friedel1969transition} the $d$-bonding energy of a transition metal is proportional to
\begin{equation}
g(N_d)=\frac{N_d\left(10-N_d\right)}{10},\qquad N_d={\rm VEC}-n_{sp},
\label{eq:g}
\end{equation}
where VEC is the valence electron concentration and $n_{sp}=1.75$ the number of electrons assigned to the $s$ and $p$ bands; $g=0$ for the $sp$ metals. The function is largest for a half-filled band and vanishes for an empty or full band, and $\bar g_M$ is its composition-weighted host value. An element with weaker $d$ bonding than the host has less cohesion to lose at an under-coordinated site and is expected to segregate. The value of $n_{sp}$ was chosen by cross-validation within the physical range of 1 to 2. The fitted $c_v$ is small and its confidence interval includes zero, so we regard the term as a correction.

Every descriptor is defined relative to the average host atom and vanishes for a host species in its own host. Equation~\eqref{eq:mu} therefore has no intercept and no host-specific constant, and one set of coefficients describes pure metals and concentrated alloys alike, including the redistribution of the principal elements of an HEA.

\subsection{Spectral width and species loadings}
\label{sec:width}

The loadings on the shared coordinates and the species-specific scatter are built from the same descriptors:
\begin{equation}
L_{1a}=t_w\,w_a,\qquad L_{2a}=t_s\,\Delta H^{\rm surf}_a,\qquad b_a^2=t_e^2\,E_{{\rm coh},M}\,\delta_a^2+\left(t_h\,\Delta H^\circ_{a\,{\rm in}\,M}\right)^2+s_0^2.
\label{eq:load}
\end{equation}
The strain loading is proportional to the size descriptor, so a species with a large size mismatch has a broad spectrum and responds strongly to local dilation or compression; with $t_w>0$ the loading has the sign of the misfit, consistent with the sign convention of $z_{1i}$. The bond-deficit loading is proportional to the surface-enthalpy difference: with $t_s>0$, a species of lower surface enthalpy than the host has $L_{2a}<0$ and finds its lowest energies at under-coordinated sites. The species-specific variance contains a size-mismatch part, a solution-enthalpy part and a floor $s_0=2$\kJ{}, which prevents an unphysically narrow spectrum. For the host reference itself $\mu=\sigma=0$. This construction is what makes the correlations predictable: the same size and surface descriptors that place the mean of a spectrum also fix how that spectrum moves from site to site, and therefore how it moves relative to the spectra of all other species. Table~\ref{tab:coef} lists the calibrated coefficients.

\begin{table}[t]
\centering
\caption{Calibrated coefficients of Eqs.~\eqref{eq:mu} and \eqref{eq:load}. Out-of-sample residuals: 19.0\kJ{} for $\mu$ and 0.60 for $\ln\sigma$.}
\label{tab:coef}
\footnotesize
\begin{tabularx}{\textwidth}{p{3.0cm}p{2.6cm}X}
\toprule
coefficient & value & physical interpretation\\
\midrule
$c_s$ & $+0.074$ & surface (broken-bond) contribution\\
$c_e$ & $-0.024$ & partial relaxation of the misfit strain at the boundary\\
$c_h$ & $-0.0001$ & solution-enthalpy contribution, negligible for binaries\\
$c_w$, $c_{|w|}$ & $-0.061$, $-0.159$ & signed and magnitude-dependent size contributions\\
$c_v$ & $+1.14$\kJ & $d$-band correction; confidence interval includes zero\\
$t_w$, $t_s$ & $0.221$, $0.053$ & loadings on the shared strain and bond-deficit coordinates\\
$t_e$, $t_h$, $s_0$ & $0.52$, $0.022$, 2\kJ & species-specific width and its floor\\
\bottomrule
\end{tabularx}
\end{table}

\subsection{Interactions among the segregated species}
\label{sec:inter}

The regular-solution parameter of Eq.~\eqref{eq:phi} is the mean of the two dilute solution enthalpies,
\begin{equation}
\Omega_{ab}=\tfrac12\left(\Delta H^\circ_{a\,{\rm in}\,b}+\Delta H^\circ_{b\,{\rm in}\,a}\right),
\label{eq:omega}
\end{equation}
so that the mixing enthalpy $\Omega_{ab}X(1-X)$ of an $a$-$b$ solution has the right magnitude at both dilute ends. The solution enthalpies come from Miedema's model~\cite{miedema1980cohesion,deboer1988cohesion},
\begin{equation}
\Delta H^\circ_{a\,{\rm in}\,b}=\frac{2V_a^{2/3}}{\left(n_{ws}^{-1/3}\right)_a+\left(n_{ws}^{-1/3}\right)_b}\left[-P\left(\Delta\phi^*\right)^2+Q\left(\Delta n_{ws}^{1/3}\right)^2-R'\right],
\label{eq:miedema}
\end{equation}
where $\Delta\phi^*$ is the difference between the Miedema electronegativities of $a$ and $b$, $\Delta n_{ws}^{1/3}$ the difference between the cube roots of their electron densities at the boundary of the Wigner-Seitz cell, $V_a^{2/3}$ the molar surface area of $a$, and $P$, $Q$ and $R'$ the constants of the model. The first term in the bracket is the charge-transfer attraction, the second the repulsion from the electron-density mismatch, and the third a hybridization term that applies when a transition metal is combined with a $p$-block element. Tabulated enthalpies are used where they exist \cite{deboer1988cohesion,zhang2016miedema} and Eq.~\eqref{eq:miedema} otherwise; the computed values reproduce 590 tabulated enthalpies with a mean absolute error of 11.1\kJ{} ($r=0.986$). A strongly negative $\Omega_{ab}$ marks a pair that prefers to be together, as for the compound formers Ti and P or Mn and S.

The parameter $\lambda$ scales the interaction at the boundary relative to the bulk. We use $\lambda=1$, because the available comparisons place the ratio near unity with some scatter: 0.96 for Al-Mg~\cite{matson2021atomistic}, a median of 0.78 for 17 Cu binaries in an EAM dataset~\cite{matson2024grain}, and 1.2 to 1.5 for computed metal pairs at Fe boundaries (Section~\ref{sec:res_fe}).

A second interaction acts between atoms of misfitting size. At a boundary, an oversized and an undersized atom relieve part of each other's strain, whereas two atoms that misfit in the same direction add to it:
\begin{equation}
W_{ab}=\kappa\,E_{{\rm coh},M}\,\delta_a\,\delta_b,\qquad \kappa=|c_e|=0.024.
\label{eq:Wel}
\end{equation}
Misfits of opposite sign give $W_{ab}<0$, an attraction. The term is applied to pairs that include a dilute species, since the principal elements define the average lattice. The coefficient $\kappa$ is not fitted separately; it is tied to the fraction of the misfit energy that a single solute recovers at the boundary in Eq.~\eqref{eq:mu}.

The mean-field expressions of Eq.~\eqref{eq:phi} assume that the boundary atoms are randomly mixed. A strongly attracting pair can lower its free energy further by placing the partners on neighboring sites, and we add a first-shell quasi-chemical correction for attracting pairs. Let a boundary site have $z$ neighboring boundary sites ($z=8$ in bcc and 12 in close-packed hosts), each occupied by $b$ with probability $\Xb_b$, and let $u_{ab}$ be the energy of one $a$-$b$ bond relative to the host. The free energy of the shell is $-zRT\ln[1+\Xb_b(e^{-u_{ab}/RT}-1)]$, of which the mean field keeps only the linear part $z\,u_{ab}\Xb_b$. The correction to the boundary chemical potential is the remainder,
\begin{equation}
\Delta\phi_a=\sum_b\left[-zRT\ln\!\left(1+\Xb_b\left(e^{-u_{ab}/RT}-1\right)\right)-z\,u_{ab}\,\Xb_b\right],\qquad u_{ab}=\frac{s_M\,W^{\rm M}_{ab}}{z},
\label{eq:qc}
\end{equation}
with the Miedema ternary interaction
\begin{equation}
W^{\rm M}_{ab}=\Omega_{ab}-\Omega_{aM}-\Omega_{bM},
\label{eq:WM}
\end{equation}
the energy gained when an $a$-$b$ contact replaces an $a$-$M$ and a $b$-$M$ contact. The scale $s_M$ converts it to a boundary pair energy in host $M$ and is taken from atomistic pair energies, not from the multicomponent observations: 0.26 in Mg from DS~\cite{zhang2026predicting}, 1.3 in Fe from MLIP and DFT pair energies (Section~\ref{sec:res_fe}), 0.85 in Cu and 0.4 in Al from the dataset of Matson and Schuh~\cite{matson2024grain}, and 1 in other hosts.

\subsection{Light elements on an impurity sublattice}
\label{sec:light}

Hydrogen, boron, carbon, nitrogen, oxygen and sulfur occupy the open spaces between the metal atoms of a boundary rather than its metallic sites. We place them on an impurity sublattice with $r_b$ bulk sites per metal atom and $r_{\gb}$ open boundary sites per boundary metal atom. The bulk reference sites are the lowest-energy sites of the DFT database~\cite{mai2026grain}: octahedral for C, N and O and tetrahedral for H in bcc hosts, octahedral in fcc and hcp hosts, and substitutional for B and S. Phosphorus is substitutional in bulk Fe but occupies open positions at Fe boundaries, so it is placed on the impurity sublattice with a substitutional bulk reference; P, C, B and S then compete for the same open boundary sites, as observed experimentally. We use $r_{\gb}=1$ in Fe, which corresponds to saturation of P and S at about one monolayer~\cite{erhart1981equilibrium,lejcek2017interfacial}, and $r_{\gb}=4$ in other hosts.

The occupancy of open boundary site $i$ by light element $X$ is given by the multi-species Fermi-Dirac expression
\begin{equation}
\theta_{iX}=\frac{\zeta_X\exp\!\left[-\left(E_{iX}+s_X\right)/RT\right]}{1+\sum_Y\zeta_Y\exp\!\left[-\left(E_{iY}+s_Y\right)/RT\right]},
\label{eq:light}
\end{equation}
together with the conservation of each light element,
\begin{equation}
f\,r_{\gb}\,\bar\theta_X+(1-f)\,r_b\,\theta^c_X=L_X.
\label{eq:lightmass}
\end{equation}
Here $E_{iX}$ is the segregation energy of $X$ at site $i$ relative to its bulk reference site, $s_X$ its chemical coupling to the metals defined in Eq. \eqref{eq:sX}, $\bar\theta_X$ the boundary-averaged occupancy, $\theta^c_X$ the occupancy of the bulk sites, and $L_X$ the total content of $X$ per metal atom. The activity $\zeta_X$ is adjusted until Eq.~\eqref{eq:lightmass} holds. The 1 in the denominator of Eq.~\eqref{eq:light} represents an empty site; because all light elements share one finite set of sites, and are mutually exclusive site by site. Their site energies are sampled from a joint light-element spectrum with the correlations computed in Fe (Section~\ref{sec:res_fe}).

The two sublattices are coupled chemically. The boundary energy of light element $X$ is shifted by
\begin{equation}
s_X=\eta\sum_a\left(\Xb_a-X^c_a\right)\Delta H(X\ {\rm in}\ a),
\label{eq:sX}
\end{equation}
where $\Delta H(X\ {\rm in}\ a)$ is the enthalpy of solution of $X$ in metal $a$ and $\Xb_a-X^c_a$ is the excess of $a$ at the boundary. A metal that is enriched at the boundary and binds $X$ strongly draws $X$ to the boundary; one with a positive enthalpy pushes it away. The metals feel the same interaction in return: the term $\eta\,[r_{\gb}\sum_X\bar\theta_X-r_b\sum_X\theta^c_X]\,\Delta H(X\ {\rm in}\ a)$ is added to $\phi^{\gb}_a-\phi^c_a$ in Eq.~\eqref{eq:iso}, so that one interaction energy acts on both sublattices. We use $\eta=0.18$, the slope of 11 metal-metalloid pair energies computed at Fe boundaries against their Miedema values (Section~\ref{sec:res_fe}).

In pure Fe, the light-element spectra are anchored to measured polycrystalline segregation free energies. Each spectrum is centered on $\Delta G_X-RT\ln r_b$, which converts a McLean free energy into a site energy for $r_b$ bulk sites, and given a width of 5\kJ, while the DFT site correlations are kept. We show why this is needed in Section~\ref{sec:res_fe}. For P and C we use $\Delta G_{\rm P}=-34.3\,{\rm kJ\,mol^{-1}}-21.5\,T\,{\rm J\,mol^{-1}K^{-1}}$ and $\Delta G_{\rm C}=-76$\kJ{} at 873\,K from Erhart and Grabke~\cite{erhart1981equilibrium}; for S, Briant's polycrystal value of $-51.5$\kJ{} at 823\,K with the segregation entropy of $-19.3$\,J\,mol$^{-1}$K$^{-1}$~\cite{lejcek2017interfacial,lejvcek2002prediction}; for N and B, the compiled general-boundary enthalpies with the enthalpy-entropy compensation rule of $\alpha$-Fe~\cite{lejcek2017interfacial,lejcek2016interstitial}. O and H keep their DFT effective energies. In other hosts the light-element spectra are transferred from the Fe and Ni data and are treated as less certain.

\subsection{Bulk sinks and the dissolved reservoir}
\label{sec:sinks}

Boundary segregation is controlled by the dissolved solute, not by the amount added to the alloy. A strong sulfide or phosphide former can lower the segregation of an impurity simply by binding it inside the grains. In ferrite we account for four sinks, each of which lowers $X^{\rm tot}_a$ in Eq.~\eqref{eq:mass} before the boundary equilibrium is solved. Sulfur precipitates as MnS and CrS according to the solubility products of the CALPHAD assessment of Oikawa et al.~\cite{oikawa2000thermodynamic}; other sulfides are scaled from MnS with the Miedema enthalpies. Phosphorus precipitates as TiP, whose solubility is anchored on the precipitation reported by Grabke et al.\ in Fe-0.08Ti-0.01P (wt.\%) at 1173\,K~\cite{grabke1987effects}, and as FeNbP, taken from the Fe-Nb-P assessment of Tokunaga et al.~\cite{tokunaga2009thermodynamic}; other phosphides are scaled with the Miedema formation enthalpies. Dissolved carbon is limited by the cementite solvus. Finally, attracting metal-metalloid pairs bind part of each partner in the grains. For a pair with grain concentrations $X^c_a$ and $X^c_b$, the concentration $n$ of bound pairs follows from the nearest-neighbor association relation
\begin{equation}
\frac{n}{\left(X^c_a-n\right)\left(X^c_b-n\right)}=z\left[\exp\!\left(-\frac{s\,W^{\rm M}_{ab}}{zRT}\right)-1\right],
\label{eq:pairs}
\end{equation}
with $s=0.26$ in bcc Fe, the ratio of the computed nearest-neighbor pair energies in bulk Fe to the Miedema estimate. Paired atoms are removed from the reservoir that can segregate. The parameters of all sinks are listed in Supplementary Note~S5.

\subsection{Calibration and numerical solution}
\label{sec:calib}

Only binary data are used to calibrate the theory. The calibration set contains 286 solute-host spectra in 18 hosts. Of these, 219 are the EAM polycrystal spectra of Wagih et al.~\cite{wagih2020learning}, averaged over the interatomic potentials available for each pair. In Al and Fe, 67 DFT spectra take the place of the EAM data: 31 Al solutes from the DFT-accuracy Al polycrystal of Wagih et al.~\cite{wagih2025designing}, and 36 metal solutes on the substitutional sites of six bcc Fe bicrystals from the database of Mai et al.~\cite{mai2026grain}, for which we recomputed every segregation energy relative to the relaxed pure boundary and to the lowest-energy bulk site of the element. The six mean coefficients of Eq.~\eqref{eq:mu} were fitted by ordinary least squares to the spectral means. The four width coefficients of Eq.~\eqref{eq:load} were fitted by robust least squares to $\ln\sigma$ of the same spectra together with 520 measured site-by-site correlations: 465 among the Al solutes~\cite{wagih2025designing} and 55 in Mg~\cite{zhang2026predicting}. The correlations are needed because a width alone does not reveal how much of the variance is shared with other species; Eq.~\eqref{eq:rho} does. The Fe correlations rest on only 5 to 14 substitutional sites per boundary and lowered the cross-validated score when included, so they were kept for testing.

Each model variant was scored by three cross-validation splits, leaving one solute out, leaving one host out, and random tenfold. Each split was scored at 800\,K and 1\,at.\% with four measures: the balanced accuracy of the three-class call (segregated, neutral, depleted), the balanced accuracies of the segregated and depleted calls taken separately, and the Spearman rank correlation of the solutes within each host. The composite score $Q$ is their mean, and a variant was retained only if it improved $Q$ for both unseen solutes and unseen hosts. The out-of-sample residuals of $\mu$ and $\ln\sigma$ are propagated to every prediction by resampling, which gives a probability that the species segregates.

The model inputs are tabulated elemental data: molar volume, metallic radius, cohesive energy, surface enthalpy, valence electron count, and the Miedema electronegativity, electron density and volume~\cite{deboer1988cohesion}. For a given alloy, temperature and grain size, the joint spectrum is sampled on 12,000 sites (8,000 for the design maps) with a fixed random seed. The isotherm, Eq.~\eqref{eq:iso}, and the mass balance, Eq.~\eqref{eq:mass}, are solved for the grain concentrations by damped Newton iteration in $\ln X^c_a$, together with Eqs.~\eqref{eq:light} and \eqref{eq:lightmass} when light elements are present. Because strong attractions can produce more than one self-consistent solution, the solver starts from both a dilute and a saturated boundary. It then follows the solution from the non-interacting state by continuation in the interaction strength, and keeps the solution with the lowest $G$. A coupling $C_{ab}$ is evaluated from two solutions on the same sampled sites, with $b$ present and with $b$ replaced by the host, so that it contains no noise from resampling. 

The two-coordinate structure was tested on the DFT-accuracy Al polycrystal of Wagih et al.~\cite{wagih2025designing}, in which the energies of 40 solutes are available on the same 82,182 boundary sites. The energies of each solute were standardized over the sites, and the $40\times40$ matrix of their site-by-site correlations was decomposed into principal factors. Each factor has a loading for every solute and a score for every site. To relate the site scores to boundary geometry, two quantities were computed for every boundary atom in the periodic cell: a local dilation, the mean distance to its 12 nearest neighbors, and a bond deficit, 12 minus the number of neighbors within 3.4\,\AA, a cutoff between the first (2.86\,\AA) and second (4.05\,\AA) neighbor shells of Al. Atoms inside the grains give 2.862\,\AA{} and 12.0 neighbors with these definitions. Each geometric quantity was regressed on the site scores of the first factors, and the coefficient of determination $R^2$ measures how much of it the factors contain.

None of the constants $\kappa$, $\lambda$, $s_M$, $\eta$ and $s$, nor the light-element anchors, was adjusted to the multicomponent observations used below: $\kappa$ is tied to $c_e$, $\lambda$ comes from published ratios of boundary to bulk interactions, $s_M$, $\eta$ and $s$ from computed pair energies, and the anchors from measured binary segregation.

\subsection{DFT and machine-learned-potential calculations in Fe}
\label{sec:meth_fe}

The DFT database of Mai et al.~\cite{mai2026grain} contains H, B, C, N, O, P and S on a common set of interstitial and substitutional sites at six bcc Fe boundaries, $\Sigma3(1\bar11)$, $\Sigma3(1\bar12)$, $\Sigma5(210)$, $\Sigma5(310)$, $\Sigma9(2\bar21)$ and $\Sigma11(3\bar32)$, as well as metal solutes on substitutional sites. We referenced every energy to the relaxed pure boundary and to the element's lowest-energy bulk site with the number of Fe atoms balanced; the recalculated minima agree with the published values within 0.05\,eV in 37 of 48 cases and within 0.10\,eV in 47. Because all light elements were placed on the same starting sites, the database also gives their site-by-site correlations.

To obtain energies on a realistic boundary network, we first assessed four foundation machine-learned interatomic potentials (MLIPs) by recomputing 624 segregation energies of this database (seven light elements and nine metals at the six boundaries). MACE MatPES-PBE~\cite{batatia2022mace,batatia2025foundation,kaplan2025foundational} reproduced them with a mean absolute error (MAE) of 0.13\,eV and $r=0.93$ after relaxation; the models trained on Materials Project data were off by 0.6 to 4.2\,eV, mostly because they had learned GGA$+U$ energies for Fe-O. A periodic Voronoi polycrystal of 82,699 atoms (9.9\,nm cell, 12 grains of about 5\,nm) was annealed for 50\,ps at 600\,K with the Mendelev EAM potential~\cite{mendelev2003development}, cooled, relaxed and rescaled to the MACE lattice parameter (2.8506\,\AA); 33.5\% of its atoms are not bcc-coordinated according to polyhedral template matching (PTM)~\cite{larsen2016robust}. With MACE, clusters of 14\,\AA{} radius were cut around selected sites, with atoms beyond 6\,\AA{} from the site held fixed, and relaxed with and without the solute. In total, 17,604 relaxations gave light-element energies on 405 interstitial boundary sites, energies of B, P, S and nine metals on 150 substitutional sites, and the energies of 13 metal-metalloid and eight metal-metal pairs on neighboring boundary sites and in perfect bcc Fe. Increasing the cluster radius to 17\,\AA{} changed 140 test energies by 11\,meV on average.

\subsection{Hybrid MD/MC and lattice Monte Carlo on the polycrystal}
\label{sec:meth_lmc}

For simulations of the whole polycrystal we trained a neuroevolution potential (NEP)~\cite{fan2021neuroevolution} with GPUMD~\cite{xu2025gpumd} by distillation from MACE MatPES-PBE. The training set contained 3,100 structures labeled with MACE: 900 bulk bcc Fe cells with substituted solutes, 300 solute clusters, 1,300 configurations of the six DFT bicrystals with solutes near the boundary, 300 small polycrystals and 300 competing compounds and elements (among them MnS, CrS, TiS, TiP, NbP, VP, MoP, CrP, MnP and Fe$_2$P), all with random displacements of 0.02 to 0.12\,\AA{} and strains of up to 2\%. The potential covers Fe and the 11 solutes P, S, Sn, Ti, Nb, Mo, V, Cr, Ni, Cu and Mn (NEP4, radial and angular cutoffs of 6 and 4\,\AA). It was accepted only after reproducing the 624 DFT segregation energies with a relaxed MAE below 0.2\,eV; the universal NEP89 potential failed this test (0.40\,eV, $r=0.55$).

The hybrid MD/MC simulations (Fig.~\ref{fig:ovito_bin}) were run in GPUMD on the 82,699-atom polycrystal with each solute at 0.5\,at.\% (about 410 atoms) placed on randomly chosen Fe sites. After 20\,ps of MD at 773\,K in the isothermal-isobaric ensemble (Berendsen thermostat and barostat at zero pressure, time step 1\,fs), 300\,ps of hybrid MD/MC followed, with 1,000 canonical swap trials between unlike atoms every 100 MD steps ($3\times10^6$ trials), and each run was then continued for another 300\,ps from its last frame. Boundary atoms were identified in every saved frame as the atoms that PTM does not classify as bcc, so that solute following a migrating boundary is still counted at the boundary. The boundary fraction of a solute is the fraction of its atoms on boundary sites, averaged over the second half of the frames.

The lattice MC simulations (Fig.~\ref{fig:ovito_tern}) separate the thermodynamics from the dynamics of the boundary network. With the NEP, the segregation energy of each of the 11 solutes was computed on every one of the 25,085 boundary sites of the polycrystal. Solute atoms were placed on boundary sites at least 10\,\AA{} apart, each configuration was minimized with the FIRE algorithm to a force tolerance of 5\,meV\,\AA$^{-1}$, and the energy of each site was taken as the change of the per-atom energies within 5\,\AA{} of the solute, minus the median of the same quantity at eight grain-interior reference sites at least 9\,\AA{} from any boundary atom. Between 0.1 and 1.9\% of the values per solute were outliers and were capped at five robust standard deviations. On 122 sites per solute shared with the MACE clusters, the NEP energies agree with MACE with $r=0.30$ to 0.75 and MAE 0.07 to 0.26\,eV. Pair energies for all 66 pairs of the 11 solutes were computed with MACE for first- and second-neighbor pairs in bulk bcc Fe and for neighboring boundary sites averaged over the six bicrystals. The lattice energy is
\begin{equation}
H=\sum_ih_i(s_i)+\sum_{\langle ij\rangle}J_{t(ij)}(s_i,s_j),
\label{eq:lmc}
\end{equation}
where $s_i$ is the species on site $i$, $h_i$ its segregation energy at that site (zero in the bulk and for Fe), and $J_t$ the pair energy for the bond class $t$ of the pair $\langle ij\rangle$: a boundary pair if both sites are boundary sites, and otherwise a bulk first-neighbor ($r<2.66$\,\AA) or second-neighbor ($r<3.1$\,\AA) pair. A trial move swaps a randomly chosen solute atom on site $i$ with the atom on site $j$, where $j$ is drawn with probability $q(j)=p_{\gb}/N_{\gb}+(1-p_{\gb})/N$ for boundary sites and $(1-p_{\gb})/N$ otherwise, with $p_{\gb}=0.5$, $N_{\gb}$ the number of boundary sites and $N$ the number of all sites. To keep detailed balance with these biased proposals, a solute-Fe swap is accepted with probability
\begin{equation}
a=\min\left[1,\ \frac{q(i)}{q(j)}\,e^{-\Delta H/k_BT}\right],
\label{eq:acc}
\end{equation}
where $\Delta H$ is the change of Eq.~\eqref{eq:lmc} and $k_B$ the Boltzmann constant; a solute-solute swap is accepted with $\min[1,e^{-\Delta H/k_BT}]$. Each run comprised $2\times10^7$ equilibration trials and 100 samples separated by $2\times10^5$ trials, at 773\,K with 0.5 or 3\,at.\% of each solute. Every system was run twice, from a random start and from a start with all solutes at the boundaries, and the difference gives the uncertainty. Without pair energies the sampler reproduces the exact Fermi-Dirac result (boundary fraction 0.9302 against 0.9308), and both starts converge to the same state. In the MC, the coupling is measured as the change of $\ln(\Xb_a/X^c_a)$ when the partner is added; it differs from Eq.~\eqref{eq:C} only through the small change of the grain concentration.

The predictions of the theory for the conditions of both simulations (773\,K, 0.5 and 3\,at.\% of each solute, and a boundary fraction of 0.30, which corresponds to $d=4.46$\,nm in Eq.~\eqref{eq:f}) were written to a time-stamped file before the simulations were run, together with the tests to be applied (Supplementary Note~S9). The comparison uses the predictions without bulk sinks, since the simulations contain no second phases, and the predictions were not changed afterwards.

\subsection{Multicomponent benchmark}
\label{sec:meth_bench}

The benchmark contains 137 qualitative directions taken from the literature and listed with their sources in Supplementary Tables~S5 to S9. It comprises five solute pairs in pure hosts; 12 clean equilibrium Fe ternaries, evaluated at the measured compositions and temperatures~\cite{erhart1981equilibrium,grabke1987effects,briant1980effects,tauber1978grain}; the 29 ternary interactions compiled by Xing et al.~\cite{xing2018solute}, evaluated at 773\,K and 50\,$\mu$m grains; two multi-solute co-segregation cases; 23 host-element directions in six HEAs~\cite{li2020grain,mccarthy2021emergence,zhu2024model}; ten enrichment rankings and single-solute directions, six of them in $\gamma$-Fe~\cite{darling2011stabilized,ito2022first}; and 56 Mg directions from Zhang and Deng~\cite{zhang2026predicting}. Compositions given in wt.\% were converted to atom fractions. A segregation direction is scored with the thresholds on $\beta_a$ given above, a coupling direction with $C_{ab}$ of Eq.~\eqref{eq:C}, where $|C_{ab}|<0.2$ counts as no effect, and host-element directions with the median of three independent site samples. Grain sizes of 30\,nm were used for nanocrystalline films and 50\,$\mu$m for coarse-grained material. Each case also carries a confidence grade according to whether the cited work verifies the direction directly, rests on a single simulation or on indirect or non-equilibrium evidence, or is a classical or review statement whose primary measurement we could not check. For the entries compiled by Xing et al., the grade follows the references of their Table~1. For the ablation, the ingredients of the theory were added one at a time. The baseline with independent spectra keeps the mean and width of every species but removes the shared coordinates and all interactions, which is the dilute multicomponent spectral treatment~\cite{marchiy2026multicomponent} evaluated with our predicted spectra.

\section{Results and discussion}
\label{sec:results}

\subsection{Segregation spectra of unseen hosts from elemental data}
\label{sec:res_bin}

\begin{figure}[!ht]
\centering
\includegraphics[width=\linewidth]{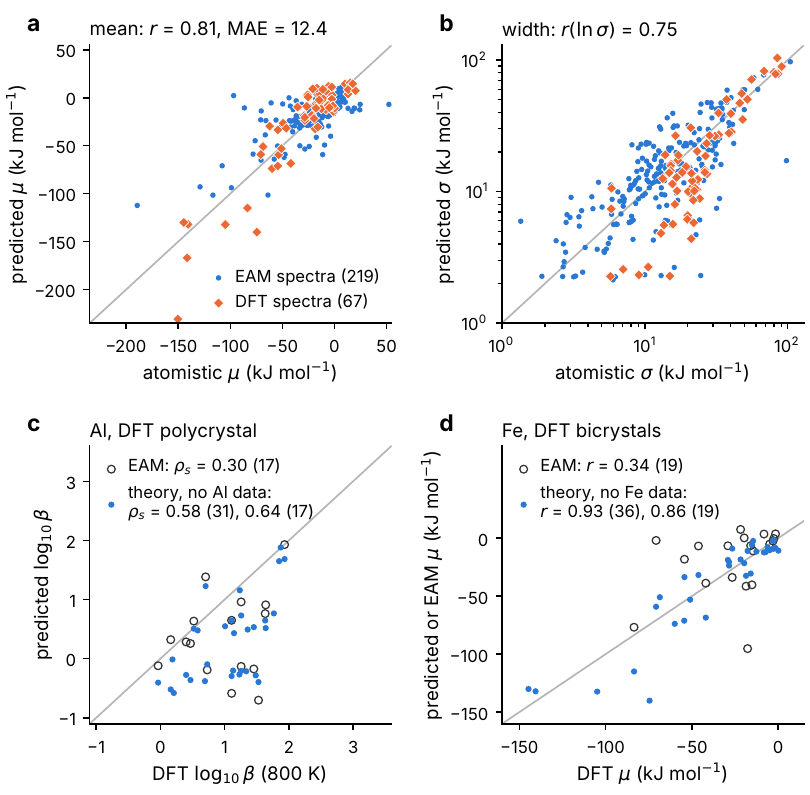}
\caption{Binary spectra predicted from elemental data. (a, b) Spectral mean and width predicted for hosts left out of the calibration, against the atomistic values: EAM spectra of Wagih et al.~\cite{wagih2020learning} (circles) and DFT spectra in Al and Fe (diamonds). (c) Al: boundary enrichment at 800\,K and 1\,at.\% in the DFT polycrystal~\cite{wagih2025designing}, against the theory calibrated without any Al data (filled) and against the EAM spectra (open). (d) Fe: DFT spectral means on six bicrystals~\cite{mai2026grain}, against the theory calibrated without any Fe data (filled) and against the EAM spectra (open). Values in brackets refer to the solutes common to both comparisons.}
\label{fig:binaries}
\end{figure}

The first question is whether tabulated elemental properties are enough to predict the spectra of a host the theory has never seen. Figures~\ref{fig:binaries}a and b answer it with leave-one-host-out predictions, in which every host is removed from the calibration in turn and its spectra are predicted from the remaining 17. The predicted spectral means of the 286 binaries correlate with the atomistic values with $r=0.81$ and an MAE of 12.4\kJ{} over a range of about 250\kJ, and the widths correlate with $r=0.75$ in $\ln\sigma$. None of these predictions used the spectra, the interatomic potential or the boundary structure of the host in question. The qualitative scores in Table~\ref{tab:val} are well above chance for unseen solutes and unseen hosts alike. Gradient-boosted trees and random forests trained on the same descriptors scored lower out of sample, indicating that the physically signed linear form generalizes better than a flexible regressor on data of this size.

\begin{table}[t]
\centering
\caption{Out-of-sample qualitative scores for binaries at 800\,K and 1\,at.\% (balanced accuracies; chance is 0.33 for three classes and 0.5 for two).}
\label{tab:val}
\footnotesize\setlength{\tabcolsep}{4pt}
\begin{tabular}{lcccccc}
\toprule
split & 3-class & segregated & depleted & rank $\rho_s$ & $Q$ & $\mu$ MAE\\
 & & & & & & (kJ\,mol$^{-1}$)\\
\midrule
unseen solute & 0.56 & 0.74 & 0.67 & 0.56 & 0.630 & 12.3\\
unseen host & 0.55 & 0.72 & 0.66 & 0.59 & 0.631 & 12.4\\
random tenfold & 0.51 & 0.73 & 0.59 & 0.56 & 0.600 & 11.8\\
\bottomrule
\end{tabular}
\end{table}

The more demanding test is against DFT, because a theory fitted largely to EAM spectra might be expected to inherit the errors of the EAM potentials. Figures~\ref{fig:binaries}c and d show that it does not. With all Al data removed from the calibration, the theory ranks the DFT enrichment of 31 Al solutes with Spearman $\rho_s=0.58$; for the 17 solutes that also have EAM spectra it reaches 0.64, whereas the EAM spectra of the same solutes reach 0.30. In bcc Fe, with all Fe data removed, the predicted means correlate with the DFT values of 36 solutes with $r=0.93$; for the 19 solutes with EAM spectra the theory gives $r=0.86$ and the EAM spectra $r=0.34$. A theory trained on many approximate spectra therefore comes closer to DFT than the individual atomistic spectra do. The errors of different potentials are largely unrelated and average out across 18 hosts and many potentials, while the descriptors carry the physics that the potentials share. The single available Fe-Mn potential illustrates the point: it gives $\mu=-40$\kJ{} with $\sigma=61$\kJ, which would make Mn a strong segregant in Fe, and the theory does not repeat this.

The main systematic error appears in Al. The late transition metals Ni, Co, Ru, Rh, Pd, Os, Ir and Pt segregate in the DFT polycrystal, with $\log_{10}\beta$ between 1 and 1.6, but are predicted to be nearly neutral; they form the group below the diagonal in Fig.~\ref{fig:binaries}c. Descriptors calibrated mostly on transition-metal hosts miss how a late transition metal bonds in an $sp$-metal host. A correction fitted to Al alone could not be validated in a second $sp$ host, so none was included.

\subsection{Two shared coordinates organize the joint spectrum}
\label{sec:res_joint}

\begin{figure}[!ht]
\centering
\includegraphics[width=\linewidth]{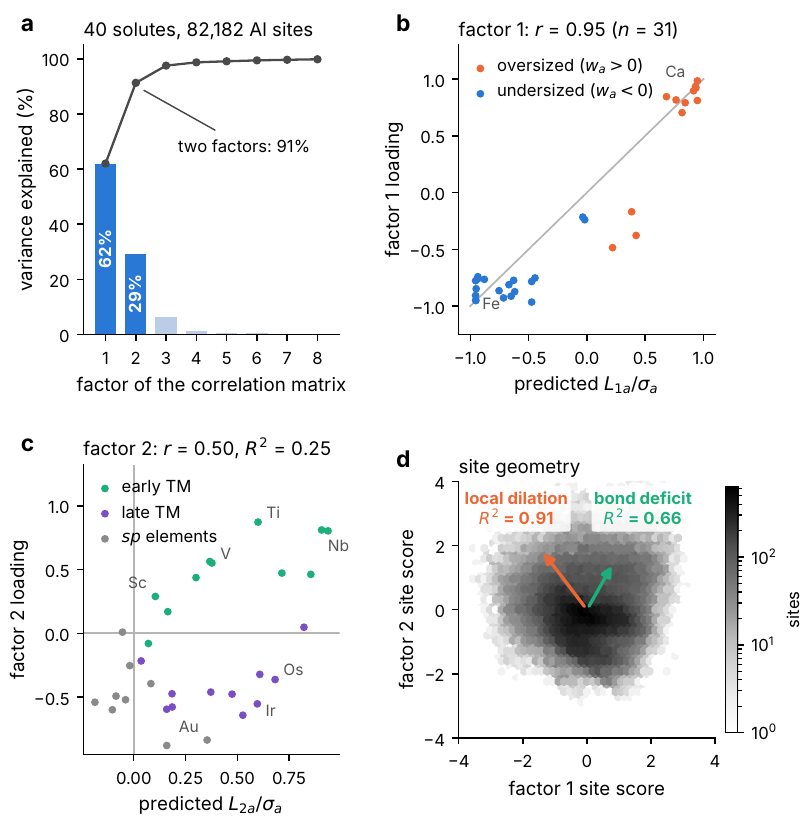}
\caption{The two-coordinate structure of boundary site energies in the DFT Al polycrystal~\cite{wagih2025designing}. (a) Variance carried by each factor of the $40\times40$ matrix of site-by-site correlations of all 40 solutes. (b, c) Loadings of the first and second factors against the normalized loadings $L_{1a}/\sigma_a$ and $L_{2a}/\sigma_a$ of the calibrated theory, for the 31 solutes with tabulated elemental properties. (d) Site scores of the 82,182 boundary sites on the two factors, with the directions along which local dilation (mean distance to the 12 nearest neighbors) and bond deficit (12 minus the number of neighbors within 3.4\,\AA) increase; arrow lengths scale with $\sqrt{R^2}$, the fraction of each geometric quantity reproduced by the two factors.}
\label{fig:twocoord}
\end{figure}

The DFT-accuracy Al polycrystal of Wagih et al.~\cite{wagih2025designing} gives the energies of 40 solutes on the same 82,182 boundary sites and therefore a direct test of the two-coordinate picture (Section~\ref{sec:calib}). The first factor of the $40\times40$ matrix of site-by-site correlations carries 62\% of the variance and the second another 29\%, while the third adds only 6\% (Fig.~\ref{fig:twocoord}a). Two factors therefore account for 91\% of the co-variation of 40 chemically very different solutes.

Both factors have a physical meaning, on the side of the solutes and on the side of the sites. On the side of the solutes, the factor loadings can be compared directly with the normalized loadings $L_{1a}/\sigma_a$ and $L_{2a}/\sigma_a$ of the calibrated theory, which are the quantities that enter Eq.~\eqref{eq:rho}. The first factor follows $L_{1a}/\sigma_a$ with $r=0.95$ for the 31 solutes with tabulated elemental data (Fig.~\ref{fig:twocoord}b). This is more than a separation of oversized from undersized solutes, because within each of the two groups the loadings still follow the size descriptor $w_a$, with $r=0.62$ and $0.55$. The second factor is reproduced less well, with $r=0.50$ and $R^2=0.25$ (Fig.~\ref{fig:twocoord}c), and the residual is systematic: the early transition metals Sc, Ti, V and Nb lie above the prediction and the late transition metals Os, Ir, Pt and Au below it. The surface enthalpy therefore misses a $d$-band contribution to the response to under-coordinated sites, the same deficiency that leaves the segregation of late transition metals in Al underestimated (Section~\ref{sec:res_bin}).

On the side of the sites, the two factors reproduce the local dilation of the 82,182 boundary sites with $R^2=0.91$ and their bond deficit with $R^2=0.66$, and adding four further factors raises these values only to 0.93 and 0.72. Neither geometric quantity coincides with one statistical factor. Dilation correlates with the first and second site scores at $r=-0.58$ and $0.76$, and bond deficit at $0.40$ and $0.70$; the two quantities are themselves correlated at boundary sites ($r=0.33$) and point about $67^\circ$ apart in the plane of the two factors (Fig.~\ref{fig:twocoord}d). The two coordinates of Eq.~\eqref{eq:spec} are therefore an orthogonal basis of a plane spanned by local strain and bond deficit, rather than these two quantities individually. This is all the theory requires, because Eqs.~\eqref{eq:sigma} and \eqref{eq:rho} are unchanged by a rotation of the basis. It is the structural result on which the theory rests: the competition between species at a polycrystalline boundary is low-dimensional, its dimensions are geometric, and the response of each species to them is set mainly by its size and, less precisely, by its surface enthalpy.

\begin{figure}[!ht]
\centering
\includegraphics[width=0.8\linewidth]{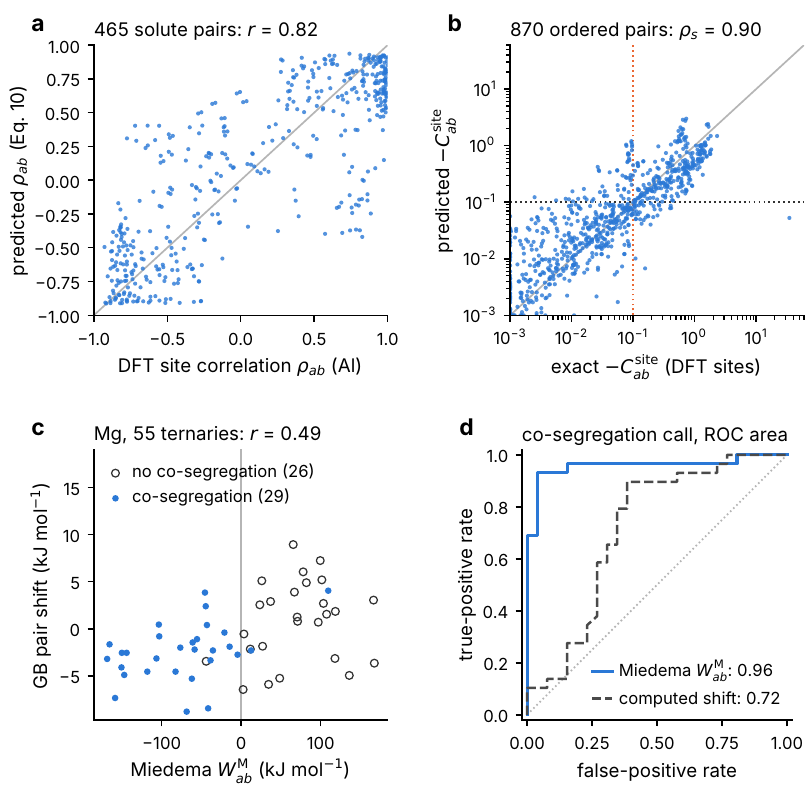}
\caption{The joint spectrum. (a) Site-by-site correlations of 465 solute pairs in the DFT Al polycrystal~\cite{wagih2025designing}, measured against predicted. (b) Site competition $-C^{\rm site}_{ab}$ for all 870 ordered Al solute pairs (1\,at.\% each, 573\,K, 50\,nm grains), computed exactly on the DFT site energies and with the model's joint spectrum. Pairs with $C^{\rm site}\ge0$, which use complementary sites (53 in the exact calculation), are plotted at the lower limit; the dotted lines mark $C=-0.1$. (c) Mg: shift of a solute's boundary spectrum when its neighbor is another solute instead of Mg, from the DS of Zhang and Deng~\cite{zhang2026predicting}, against the Miedema ternary interaction $W^{\rm M}_{ab}$. Filled symbols: the 29 pairs that co-segregate in their hybrid MD/MC. (d) Receiver operating characteristic of the co-segregation call for the same 55 pairs, with $-W^{\rm M}_{ab}$ or the negative computed shift as the score; the legend gives the area under each curve and the dotted line is a random call.}
\label{fig:joint}
\end{figure}

The theory predicts the correlations from the loadings of Eq.~\eqref{eq:load}. It reproduces the 465 DFT correlations among the Al solutes with elemental data with $r=0.82$ (Fig.~\ref{fig:joint}a), and refitting without each solute in turn gives $r=0.81$ for that solute's correlations, so the correlations of a solute can be predicted without having seen any of them. Including the correlations in the fit also improves the binary predictions: $Q$ rises from 0.541 to 0.630 for unseen solutes and from 0.555 to 0.631 for unseen hosts. The pattern is easy to read. Strongly oversized solutes share the dilated environments, and oversized and undersized solutes are anticorrelated through the strain coordinate.

A stricter test is the competition itself, since this is what the correlations are for. For all 870 ordered pairs of the 30 Al solutes we computed the site contribution to the coupling twice, once exactly on the DFT site energies of the two solutes and once on the Gaussian joint spectrum of the theory (Fig.~\ref{fig:joint}b). On the DFT sites, 34\% of the pairs compete appreciably ($C^{\rm site}<-0.1$) and only 6\% use complementary sites ($C^{\rm site}>0$). The theory ranks the exact couplings with $\rho_s=0.90$ and identifies the strongly competing pairs with a balanced accuracy of 0.88. The extreme case is Sn, which Pb displaces almost completely ($C\approx-35$). Pairwise site competition, which until now required the spectra of both partners on the same sites, can therefore be predicted from size and surface descriptors alone.

In Mg the 55 site correlations of the calibration are reproduced with $r=0.99$. The DS of Zhang and Deng add information that correlations do not contain: the shift of a solute's boundary energy when its nearest neighbor is another solute rather than Mg. This shift correlates with the Miedema ternary interaction $W^{\rm M}_{ab}$ with $r=0.49$ (Fig.~\ref{fig:joint}c). More importantly, $W^{\rm M}_{ab}$ on its own separates the 29 pairs that co-segregate in the MD/MC from the 26 that do not with an area under the receiver operating characteristic (ROC) curve of 0.96, whereas the computed shift reaches only 0.72 (Fig.~\ref{fig:joint}d). The macroscopic-atom interaction thus carries information about pair association that a single local energy shift does not, presumably because it reflects the bonding of the pair with its whole environment rather than in one configuration. Site correlations and chemical interactions are therefore two separate ingredients, and this is shown in Section~\ref{sec:res_bench} which of them decides what.

\subsection{Light elements, chemical coupling and bulk sinks in Fe}
\label{sec:res_fe}

The DFT database gives spectra for the light elements at six Fe boundaries together with their site-by-site correlations (Supplementary Tables~S2 and S3). H, B, C, N and O share interstitial sites strongly, with a correlation of 0.90 between C and N; P and S correlate with each other at 0.70 but only at 0.25 to 0.49 with the interstitial formers. The MLIP calculations on 300 common sites of the polycrystal recover the same pattern ($r=0.76$) with a higher mean correlation (0.78 against 0.56), and the theory uses the polycrystal matrix.

\begin{figure}[!ht]
\centering
\includegraphics[width=0.8\linewidth]{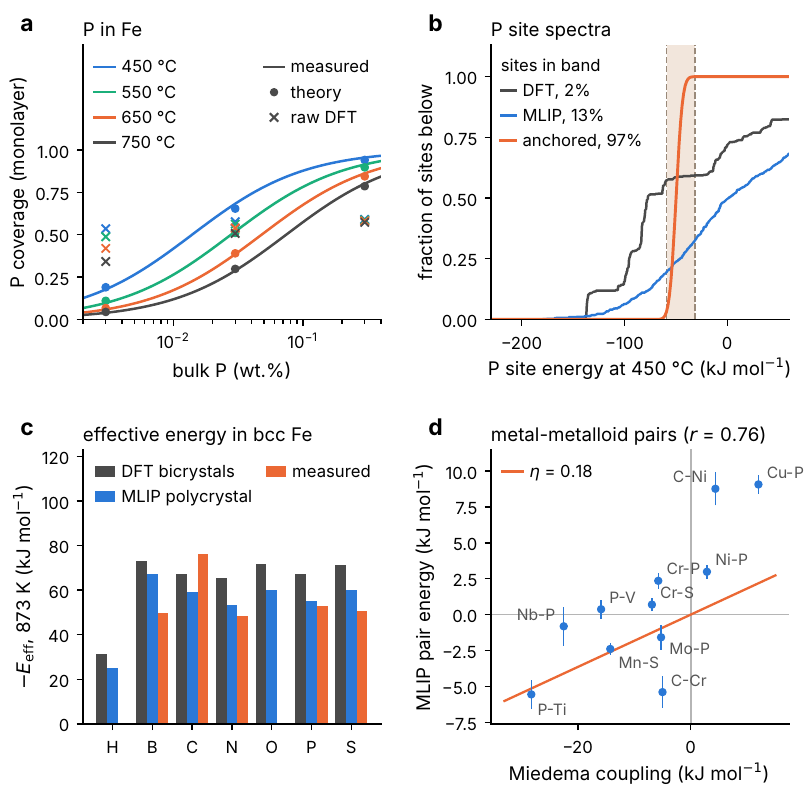}
\caption{Light elements in bcc Fe. (a) P coverage of the boundaries against bulk P content: McLean fit of Erhart and Grabke to polycrystal Auger data~\cite{erhart1981equilibrium} (lines), theory with anchored spectra (circles), and the DFT bicrystal spectrum used directly (crosses). (b) Fraction of P sites below a given energy at 450\,$^\circ$C in the DFT bicrystal spectrum (six boundaries weighted equally), the MLIP polycrystal spectrum and the anchored spectrum. A site is half filled when its energy equals $RT\ln x$, so raising the bulk content from 0.003 to 0.3\,wt.\% P fills the sites in the shaded band; the legend gives the share of each spectrum's sites that lie in it. (c) Effective segregation energy at 873\,K from the DFT bicrystals, the MLIP polycrystal, and measurements or compilations~\cite{erhart1981equilibrium,lejcek2017interfacial}. (d) Metal-metalloid pair energies on neighboring boundary sites of the polycrystal (MLIP; bars are standard errors over 100 site pairs) against the Miedema coupling per neighbor; the slope is $\eta$ of Eq.~\eqref{eq:sX}.}
\label{fig:light}
\end{figure}

Used directly, however, the broad bicrystal spectrum does not reproduce how the P coverage of Fe boundaries depends on the bulk P content (Fig.~\ref{fig:light}a). Between 0.003 and 0.3\,wt.\% P, the measured coverage of polycrystal boundaries at 450\,$^\circ$C rises from 0.18 to 0.96 of a monolayer~\cite{erhart1981equilibrium}. The unanchored DFT spectrum predicts 0.54 to 0.59 over the same range, with an RMS deviation of 0.28 over the whole dataset, and the MLIP polycrystal spectrum is equally insensitive, rising only from 0.20 to 0.33. The reason is the shape of a broad spectrum, here 40 to 90\kJ{} wide: it contains a fixed fraction of very deep sites that fill at any content and too few sites of intermediate depth to produce the gradual filling seen experimentally. Only 2\% of the DFT sites and 13\% of the MLIP sites lie in the range of energies that fills between 0.003 and 0.3\,wt.\% P, against 97\% of the anchored spectrum (Fig.~\ref{fig:light}b). This matters for multicomponent behavior, because scavenging and site competition both act through the dissolved content, and if the coverage does not respond to that content, neither effect can appear. Real boundaries keep taking up P until a monolayer is complete, probably through local restructuring that a static site spectrum does not contain. Centering the spectrum on the measured polycrystal free energy while keeping the DFT site correlations reduces the RMS deviation to 0.014 and reproduces the measured isotherm. We regard this anchoring as a calibration step for light elements in Fe and keep it in mind when transferring the light-element spectra to other hosts.

The effective segregation energies from the MLIP polycrystal are 6 to 12\kJ{} shallower than the DFT bicrystal values for every light element, and for P they lie within 2\kJ{} of the Erhart-Grabke value (Fig.~\ref{fig:light}c). The polycrystal calculations also show that strong compound formers bind P much more weakly at a boundary than the bulk Miedema enthalpies suggest. The boundary pair energies for Ti-P, Nb-P and V-P are $-5.6$, $-0.8$ and $+0.4$\kJ{} per pair, against Miedema values of $-28$, $-23$ and $-16$\kJ. Across 11 metal-metalloid pairs the computed boundary interactions follow the Miedema coupling with a slope of $\eta=0.18$ ($r=0.76$; Fig.~\ref{fig:light}d); in the bulk the same pairs bind at about 0.26 of the Miedema estimate. Neighboring light elements repel each other at the boundary, by $+11$\kJ{} for P-P and N-P and $+9$\kJ{} for C-P, which gives C and N a second way of lowering the P coverage besides excluding it from shared sites. For metal-metal pairs at Fe boundaries, the computed interactions follow about $1.5\,W^{\rm M}_{ab}/z$ ($r=0.78$ for eight pairs), and the DFT data of Mai et al.\ for 45 transition-metal pairs at four Fe boundaries~\cite{mai2022segregation} follow about $1.2\,W^{\rm M}_{ab}/z$ ($r=0.67$). Both support $\lambda$ close to one and $s_{\rm Fe}=1.3$.

\begin{table}[t]
\centering
\caption{Clean equilibrium Fe ternaries: effect of adding $M$ on the boundary content of $X$, measured and predicted at the experimental conditions. $C_{XM}$ is the coupling of Eq.~\eqref{eq:C}; $|C|<0.2$ counts as no effect.}
\label{tab:clean}
\footnotesize
\begin{tabular}{lcccl}
\toprule
effect of $M$ on $X$ & measured & predicted & $C_{XM}$ & source\\
\midrule
C on P (550\,$^\circ$C) & $-$ & $-$ & $-1.23$ & Erhart and Grabke~\cite{erhart1981equilibrium}\\
Cr on P (550\,$^\circ$C) & 0 & 0 & $+0.02$ & Erhart and Grabke~\cite{erhart1981equilibrium}\\
Ti on P (550\,$^\circ$C) & $-$ & $-$ & $-1.31$ & Grabke et al.~\cite{grabke1987effects}\\
Nb on P (550\,$^\circ$C) & $-$ & $-$ & $-1.46$ & Grabke et al.~\cite{grabke1987effects}\\
Mo on P (500\,$^\circ$C) & 0 & 0 & $+0.03$ & Grabke et al.~\cite{grabke1987effects}\\
V on P (500\,$^\circ$C) & 0 & 0 & $+0.03$ & Grabke et al.~\cite{grabke1987effects}\\
P on S (560\,$^\circ$C) & $-$ & $-$ & $-1.20$ & Briant~\cite{briant1980effects}\\
S on P (560\,$^\circ$C) & $-$ & 0 & $-0.04$ & Briant~\cite{briant1980effects}\\
Mn on S (560\,$^\circ$C) & $-$ & $-$ & $-13.5$ & Briant~\cite{briant1980effects}\\
Cr on S (560\,$^\circ$C) & $-$ & $-$ & $-4.31$ & Briant~\cite{briant1980effects}\\
N on S (600\,$^\circ$C) & $-$ & $-$ & $-0.76$ & Tauber and Grabke~\cite{tauber1978grain}\\
C on S (600\,$^\circ$C) & $-$ & $-$ & $-2.25$ & Tauber and Grabke~\cite{tauber1978grain}\\
\bottomrule
\end{tabular}
\end{table}

With these ingredients the theory reproduces 11 of the 12 clean equilibrium Fe ternaries, measured on laboratory alloys free of other additions (Table~\ref{tab:clean}). It captures the displacement of P and S by carbon, of S by P and of S by N, which come from site exclusion and repulsion on the impurity sublattice; the removal of S by MnS and CrS and of P by TiP and FeNbP, which come from the bulk sinks; and the absence of any effect of Cr, Mo and V on P, which requires the reduced metal-light coupling $\eta=0.18$, since with the full Miedema coupling these elements would drag P strongly to the boundary. Each of the three mechanisms is therefore needed, and each is supplied by a different part of the theory. The one miss is the reported lowering of P by S~\cite{briant1980effects}. With the measured polycrystal S energy, S segregates slightly less than P at 833\,K, and 30\,ppm S cannot displace 0.03\,wt.\% P; with the DFT bicrystal S energy the case is reproduced, so the prediction is sensitive to the S reference.

\subsection{Multicomponent benchmark and ablation}
\label{sec:res_bench}

Outside Mg, the full theory reproduces 58 of the 81 literature directions of the benchmark, and it reproduces 50 of the 56 Mg directions, which consist of the 55 ternaries (49 correct) and the displacement of Al by Zn. Of the 27 directions outside Mg that rest on a direct equilibrium measurement or calculation (grade A), it reproduces 24; of the other 54, which rest on commercial steels, review statements, single simulations or other indirect evidence, it reproduces 34. The ablation in Table~\ref{tab:ablation} shows how this comes about. Independent spectra, the dilute multicomponent spectral treatment with our predicted spectra, already reproduce most single-species trends, such as the host elements of HEAs, the rankings and the $\gamma$-Fe directions. They miss nearly every effect of one solute on another, with 7 of the 29 Xing directions and none of the 29 co-segregating Mg pairs, because without interactions the only effect a second solute can have is to compete. Adding the joint spectrum changes the total by one case: the correlations decide how strongly two species compete, as Fig.~\ref{fig:joint}b shows, but in the measured systems they rarely decide the sign of the effect. The chemical interactions are what turn competition into co-segregation; they raise the Xing score from 7 to 14 and identify 23 of the 29 co-segregating Mg pairs. Anchoring the Fe light-element spectra adds two cases, the bulk sinks add eight Fe directions, and the first-shell correction adds one. None of these steps involved fitting a coefficient to a multicomponent observation.

\begin{table}[t]
\centering
\caption{Ablation on the multicomponent benchmark: number of literature directions reproduced as the ingredients of the theory are added in turn. A: independent spectra with the predicted means and widths, no shared coordinates and no interactions, equivalent to the dilute multicomponent spectral treatment~\cite{marchiy2026multicomponent}. B: joint spectrum with shared coordinates. C: Miedema chemical, elastic and metal-light interactions. D: light-element spectra in Fe anchored to measured polycrystal data. E: bulk sinks. F: first-shell quasi-chemical correction, which is the full theory. The raw DFT light-element spectra are used in A to C.}
\label{tab:ablation}
\footnotesize\setlength{\tabcolsep}{3.5pt}
\begin{tabular}{lcccccccc}
\toprule
 & pairs & clean & Xing & co-seg. & HEA & rank, & total & Mg\\
 & & Fe & & & host & $\gamma$-Fe & & \\
 & (5) & (12) & (29) & (2) & (23) & (10) & (81) & (56)\\
\midrule
A: independent spectra & 0 & 5 & 7 & 2 & 18 & 7 & 39 & 27\\
B: + joint spectrum & 0 & 5 & 7 & 2 & 17 & 7 & 38 & 27\\
C: + interactions & 1 & 5 & 14 & 2 & 18 & 7 & 47 & 50\\
D: + anchoring & 1 & 7 & 14 & 2 & 18 & 7 & 49 & 50\\
E: + bulk sinks & 1 & 11 & 18 & 2 & 18 & 7 & 57 & 50\\
F: + first-shell correction & 1 & 11 & 19 & 2 & 18 & 7 & \textbf{58} & \textbf{50}\\
\bottomrule
\end{tabular}
\end{table}

The ablation thus establishes a hierarchy. Site competition is what happens by default when several species share a finite set of non-equivalent sites, and the joint spectrum sets its strength. Co-segregation requires a chemical attraction strong enough to overcome that competition, and for metal pairs the Miedema ternary interaction is a good predictor of it. For impurities in steels, the dissolved reservoir often decides the outcome before the boundary is involved at all. A model containing only site competition and boundary attraction would therefore give the wrong direction for exactly the additions that metallurgists use to control impurities.

Of the 29 ternary interactions compiled by Xing et al.~\cite{xing2018solute}, the theory reproduces 19. Several of the metal-P entries in that compilation come from commercial steels in which C, Mn and carbides also act, so where a clean ternary study exists we checked the label against it. Fe-Mo-P then becomes ``no effect''~\cite{grabke1987effects}, and Fe-V-P and Fe-Cr-P were left unscored because the primary studies disagree; on the resulting 27 cases the theory gets 19 right. The remaining misses are the co-segregation of Ni and Cu with P, to which Section~\ref{sec:res_lmc} returns, of Pb with Fe in Ni and of Sn with Cr in Fe, a weak phosphide former (W), a case in which site competition outweighs a chemical attraction (Cr-Zr), an interaction predicted where none was measured (Ni-C), and Fe-Mo-P, where the compiled measure also contains the unmeasured effect of P on Mo.

Because host elements and solutes share one reference, the same calculation that gives the segregation of a dilute addition also gives the redistribution of the principal elements. For the host elements of HEAs, the theory reproduces all five directions measured by atom-probe tomography in CoCrFeMnNi~\cite{li2020grain}, with Ni and Mn enriched and Fe, Co and Cr depleted, and 18 of 23 directions across six HEAs; most of the misses concern labels from a single-boundary simulation with one EAM potential~\cite{mccarthy2021emergence} or from an empirical valence-electron rule rather than measured segregation~\cite{zhu2024model}. In Al-Ni-Hf both solutes co-segregate, as observed~\cite{wagih2025designing}. In Mg the theory finds 23 of the 29 co-segregating pairs and correctly predicts no co-segregation for the other 26. Four of the six missed Mg pairs (Al-Cu, Al-Pd, Ca-Cu and Ni-Zn) combine a weak Miedema attraction with strong site competition and co-segregate in the simulations as clusters, a form of short-range order that a mean field with a first-shell correction cannot fully represent; for the other two (Ca-Ni and Pb-Pd) the Miedema interaction is repulsive and the computed heteroatomic shift equals the homoatomic one, so their co-segregation is not carried by a pair energy at all. In paramagnetic $\gamma$-Fe the theory reproduces four of the six directions computed by Ito and Sawada~\cite{ito2022first}; it misses Mn and Cr, which are depleted in $\gamma$-Fe but segregate in the ferromagnetic bcc Fe data used for pure Fe, because the theory has no magnetic term.

\subsection{Where atomistic resolution is required: simulations of an 82,699-atom polycrystal}
\label{sec:res_lmc}

The atomistic simulations test the mechanisms of the theory on a realistic boundary network with energies that have nothing to do with the macroscopic-atom model. They are not a strict test of every number, since the hybrid MD/MC samples only 600\,ps, the lattice MC keeps the approximations of a lattice model, and neither contains second phases.

\begin{figure}[!ht]
\centering
\includegraphics[width=0.8\linewidth]{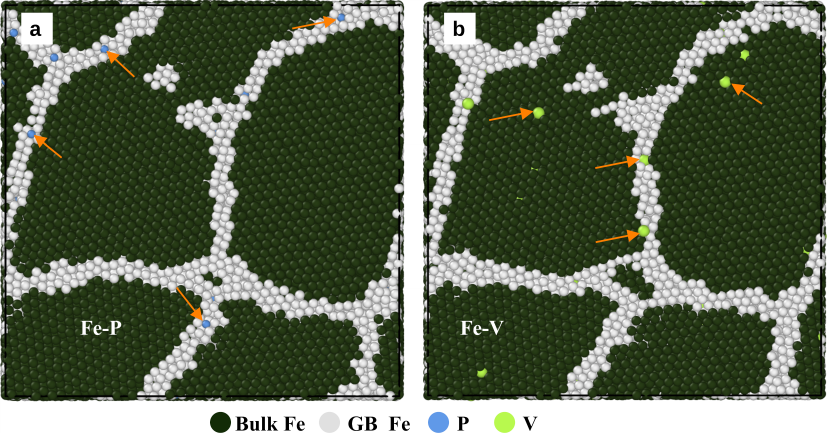}
\caption{Boundary segregation in the 82,699-atom bcc Fe polycrystal from hybrid MD/MC with the NEP potential (773\,K, 0.5\,at.\% solute). (a) Fe-P: all P atoms are at the boundaries (boundary fraction 1.00; theory 1.00). (b) Fe-V: V is shared between boundaries and grains (boundary fraction 0.44; theory 0.44). Boundary atoms, identified by polyhedral template matching as not bcc-coordinated, are shown in light grey and grain-interior Fe in dark green; arrows mark representative solute atoms.}
\label{fig:ovito_bin}
\end{figure}

At 773\,K and 0.5\,at.\%, hybrid MD/MC puts all of the P, S, Sn and Nb at the boundaries (Fig.~\ref{fig:ovito_bin}). V has a boundary fraction of 0.44, Mn of 0.75 and Cr of 0.89. The recorded predictions were 1.00 for P, S, Sn and Nb, 0.44 for V, 0.995 for Mn and 0.61 for Cr. The theory gets the complete segregation of the first group and the partial segregation of V right, overestimates Mn and underestimates Cr, two elements whose segregation in Fe is known to depend on magnetism~\cite{ito2022first}.

\begin{figure}[!ht]
\centering
\includegraphics[width=1\linewidth]{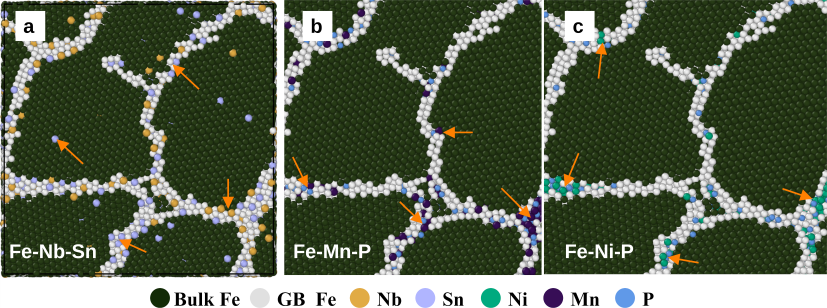}
\caption{Equilibrium lattice MC on the polycrystal with NEP site energies and MACE pair energies (773\,K, 3\,at.\% of each solute). Boundary Fe is shown in light grey and grain-interior Fe in dark green; arrows mark representative solute atoms and pairs. (a) Fe-Nb-Sn: strong competition; Nb and Sn avoid each other at the boundaries, with 0.14 times the random number of Nb-Sn neighbors. (b) Fe-Mn-P: co-segregation, with 2.85 times the random number of Mn-P neighbors. (c) Fe-Ni-P: co-segregation driven by Ni-P attraction at the boundary (3.28 times random), a case the theory misses and experiment reports.}
\label{fig:ovito_tern}
\end{figure}

\begin{figure}[!ht]
	\centering
	\includegraphics[width=1\linewidth]{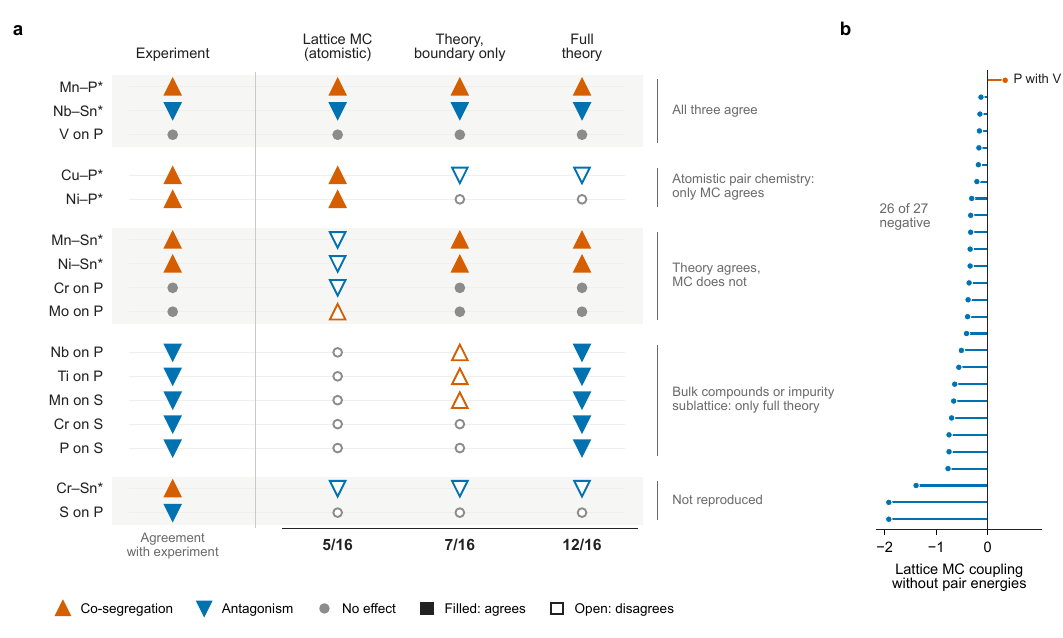}
	\caption{Measured and predicted directions for the 16 Fe pairs covered by both the lattice MC and the benchmark. (a) Experiment~\cite{xing2018solute,erhart1981equilibrium,grabke1987effects,briant1980effects}, lattice MC (3\,at.\%, 773\,K), the theory under the same boundary-only conditions, and the full theory at the experimental conditions. Asterisked pairs (Xing et al.) are scored by $C_{ab}+C_{ba}$, the others by the effect of the first element on the second. Outlined cells agree with experiment. Orange: co-segregation; blue: antagonism; grey: no effect. (b) Lattice MC couplings without pair energies, sorted.}
	\label{fig:signgrid}
\end{figure}

The ternary lattice MC supports site competition as the baseline (Fig.~\ref{fig:ovito_tern}). Without pair energies, 26 of the 27 measurable cases show a negative coupling between $-0.13$ and $-1.92$; the exception is P in the presence of V ($C=+0.35$) (Fig.~\ref{fig:signgrid}b). This is the behavior the joint spectrum predicts when the spectra of the partners are positively correlated, obtained here from site energies computed independently on 25,085 sites.

With the pair energies included, the test specified before the simulations compares the sign of each coupling predicted by the theory with the sign found in the MC. The theory agrees with the MC in 5 of the 14 measurable cases at 0.5\,at.\% and in 9 of the 30 cases at 3\,at.\%, which is close to what three equally likely outcomes would give by chance; where both calculations predict an effect, the sign agrees in 5 of 12 cases (Supplementary Note~S8). On this test, the theory and the atomistic MC disagree more often than they agree. They agree where the pair energies are strong: Nb and Sn compete strongly ($C=-4.8$ and $-5.0$ in the MC, $-1.6$ and $-1.4$ in the theory), and both calculations give co-segregation of Mn and P, enrichment of V by P, and a near-neutral P-S response.

Disagreement with the MC is not the same as disagreement with experiment, and the decisive comparison is with the measured directions for the 16 Fe pairs that both calculations cover (Fig.~\ref{fig:signgrid}a). Each calculation is scored as in the benchmark. The seven pairs taken from the compilation of Xing et al.\ are symmetric classifications and are scored by $C_{ab}+C_{ba}$, and the nine clean ternaries are directional measurements and are scored by the effect of the added element alone. On this basis the lattice MC reproduces 5 of the 16 directions, the theory under the same boundary-only conditions 7, and the full theory at the experimental conditions 12. The cases fall into distinct groups. All three calculations agree with experiment for Mn-P, Nb-Sn and V on P. For Cu-P and Ni-P the MC is right and both versions of the theory are wrong. For Mn-Sn, Ni-Sn and the absence of any effect of Cr and Mo on P, the theory is right and the MC is wrong. For the effects of Nb and Ti on P, Mn and Cr on S, and P on S, only the full theory at the experimental conditions is right; the first four involve phosphides and sulfides that form inside the grains, which no boundary-only calculation can contain however accurate its energies are. None of the calculations reproduces Cr-Sn or the effect of S on P.

The largest disagreements locate the limits of the macroscopic-atom pair chemistry. MACE bicrystal calculations give attractive boundary interactions for Ni-P and Cu-P, of $-0.04$ to $-0.11$\,eV, and with them the MC predicts strong Ni-P co-segregation ($C=+2.7$ and $+2.4$) and moderate Cu-P co-segregation ($+0.6$ and $+0.9$), in line with the reported trends and missed by the theory. The same pairs are repulsive, however, on neighboring sites of the polycrystal ($+3.0$ and $+9.1$\kJ), and the bicrystal pair energies vary strongly between boundaries. Replacing the Miedema coupling by the computed pair energies, shrunk towards the Miedema value, corrects Fe-Ni-P and Fe-Cu-P but makes Fe-Sn-S and the clean Fe-V-P case wrong, so the benchmark score does not change (Supplementary Note~S10). MACE also makes Sn repel Mn, Ni and Cr at the boundary, contrary to the co-segregation reported experimentally and predicted by the theory. The metal-metalloid pair chemistry at specific boundaries is therefore the part of the problem that needs atomistic resolution, and it needs a broader DFT assessment rather than one set of computed pair energies. This division of labor, a transferable thermodynamic theory for the whole composition space and targeted atomistic calculations for the few pairs it flags, is the practical outcome of the comparison.

\subsection{Principal elements and dilute additions in high-entropy alloys}
\label{sec:res_pred}

\begin{figure}[!ht]
\centering
\includegraphics[width=\linewidth]{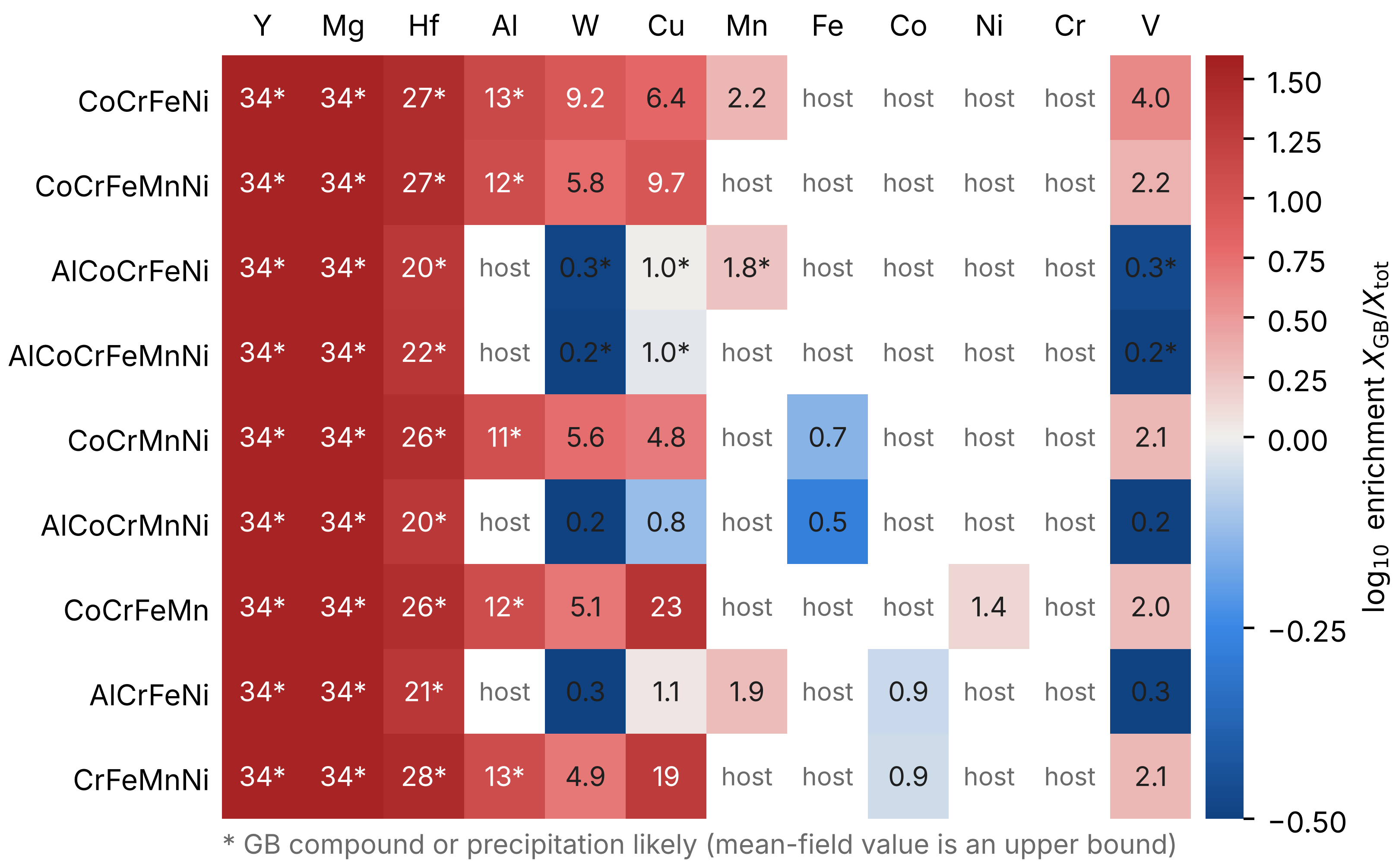}
\caption{Predicted boundary enrichment $\Xb_a/X^{\rm tot}_a$ of dilute solutes (1\,at.\%) in nine 3$d$ HEA families at 800\,K and 50\,nm grain size. Asterisks mark strongly bonded pairs concentrated at the boundary, for which a boundary compound or precipitate is likely and the mean-field value is an upper bound.}
\label{fig:hea}
\end{figure}

Because the theory needs no alloy-specific input, it can be applied across composition spaces for which no interatomic potential exists. Figure~\ref{fig:hea} shows the predicted enrichment of 12 dilute solutes in nine 3$d$ HEA families at 800\,K, 1\,at.\% solute and 50\,nm grains. Y, Mg and Hf segregate strongly in all nine. Al, W and Cu also segregate in the Al-free alloys, whereas W and V are depleted in the alloys that contain Al. Adding Al therefore changes the boundary chemistry qualitatively, removing the driving force for Cu to segregate and pushing W and V out. For strongly bonded pairs that may form a boundary compound or a precipitate, marked with asterisks, the mean-field enrichment is an upper bound rather than the composition of a stable disordered boundary.

\begin{figure}[!ht]
	\centering
	\includegraphics[width=\linewidth]{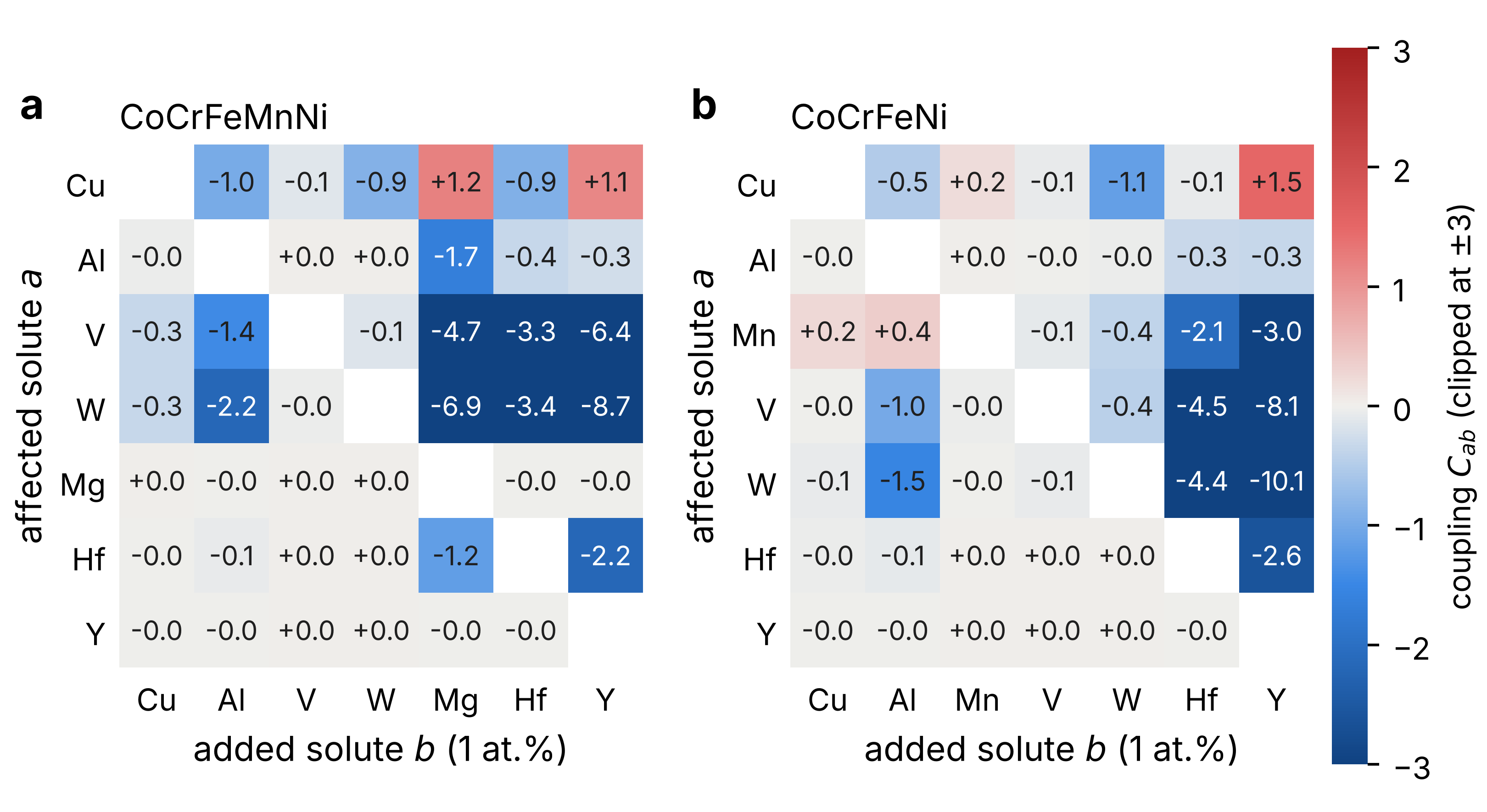}
	\caption{Predicted coupling $C_{ab}$ (Eq.~\eqref{eq:C}) between dilute solutes in CoCrFeMnNi and CoCrFeNi (1\,at.\% of each solute, 800\,K). Row: affected solute $a$; column: added solute $b$. Positive values denote co-segregation and negative values antagonism.}
	\label{fig:coupling}
\end{figure}

The predicted coupling maps (Fig.~\ref{fig:coupling}) are asymmetric, because the effect of adding $b$ on $a$ need not equal the effect of adding $a$ on $b$. In CoCrFeMnNi, Mg and Y raise the boundary concentration of Cu ($C=+1.2$ and $+1.1$), while Cu has little effect on Mg and Y, which already segregate strongly on their own. Mg, Hf and Y displace the weaker segregants V and W, with couplings as low as $-8.7$; splitting the coupling shows that most of this displacement is chemical, through the repulsion of V and W from the early metals, with an additional part from competition for dilated sites. Al does not co-segregate with any of the selected solutes but displaces V and W. Maps of this kind can be used to choose additions that steer the boundary chemistry; Y, for example, should help bring Cu to the boundaries of CoCrFeNi, while Hf or Y would keep W away. They remain predictions and should be checked where strongly bound pairs or complexions are expected. The light-element predictions for Ni, Pd and four HEAs are given in Supplementary Table~S10; in every host, B, C, N and O segregate much more strongly than H.

\subsection{Relation to atomistic spectral models}
\label{sec:discussion}

Table~\ref{tab:compare} places the theory among existing approaches. Atomistic spectral models, in particular the body of work of Wagih et al. give the most detailed picture of a given alloy: every site, every solute and, at DFT accuracy, energies that a descriptor theory cannot match site by site. Their limitation is that each alloy needs its own simulation, and the cost and the availability of potentials grow with the number of components. The present theory occupies a different position. It gives the spectrum of every species, the correlations between them, the chemical interactions, the light elements and the bulk sinks of an arbitrary alloy from tabulated data in seconds, and our results show that it does so with an accuracy that matches or exceeds that of embedded-atom spectra when both are compared with DFT. The two are complementary. The theory screens the composition space and identifies the pairs whose local chemistry it cannot resolve, as the lattice MC did for Ni-P and Cu-P, and atomistic spectra are then computed where they are needed.

\begin{table}[!ht]
\centering
\caption{Approaches to multicomponent grain-boundary segregation.}
\label{tab:compare}
\footnotesize
\begin{tabularx}{\textwidth}{>{\raggedright\arraybackslash}p{2.9cm}>{\raggedright\arraybackslash}X>{\raggedright\arraybackslash}X>{\raggedright\arraybackslash}X>{\raggedright\arraybackslash}X}
\toprule
 & single-site isotherms~\cite{mclean1957grain,fowler1939statistical,guttmann1975equilibrium,wynblatt1977surface} & atomistic spectra~\cite{wagih2020learning,wagih2022learning,wagih2025designing} & dilute multicomponent spectral isotherm~\cite{marchiy2026multicomponent} & this work\\
\midrule
\midrule
\textbf{input for a new alloy} & measured or estimated segregation energies & simulation of that alloy & simulated spectra of each solute & tabulated elemental properties\\
\midrule
\textbf{distribution of site energies} & no & computed & computed & predicted\\
\midrule
\textbf{correlation between species} & no & computed for each pair & from computed spectra & predicted from loadings\\
\midrule
\textbf{interactions among segregants} & regular solution & from additional simulations & not included & macroscopic-atom, elastic and first-shell terms\\
\midrule
\textbf{host elements of concentrated alloys} & no & in a simulation of that alloy & dilute solutes only & same isotherm as solutes\\
\midrule
\textbf{light elements with metals} & no & separate interstitial spectra & no & coupled sublattice\\
\midrule
\textbf{bulk compounds and pairs} & no & no & no & sulfides, phosphides, cementite, pairs\\
\bottomrule
\end{tabularx}
\end{table}

The reason a few descriptors suffice is the structure found in Section~\ref{sec:res_joint}. Local strain and bond deficit are not a complete microscopic description of a boundary; they are a compact basis onto which the response of each species can be projected, with atomic size and surface enthalpy fixing the projection. Because these are exactly the quantities that the macroscopic-atom model tabulates for almost every element, the theory is transferable in a way that a fitted potential is not. The common reference atom extends this transferability to concentrated alloys, in which segregation often means a redistribution among the principal elements rather than the enrichment of a trace addition.

\subsection{Scope and limitations}

The theory is a mean-field description with a first-shell correction and does not fully describe strong short-range clustering, complexion transitions or boundary reconstruction. The segregation of late transition metals in Al is systematically underestimated, which shows that the present descriptors miss part of the bonding of transition metals in $sp$-metal hosts. Except in Al and Fe, the spectral calibration rests on EAM data; averaging over hosts and potentials reduces potential-specific errors but cannot remove them. The light-element spectra are anchored mainly in Fe, and their transfer to other hosts is less certain. There is no magnetic term, so ferromagnetic $\alpha$-Fe and paramagnetic $\gamma$-Fe cannot be distinguished. Bulk sinks are implemented only for ferrite, and alloy carbides other than cementite are not yet included. Finally, the theory describes equilibrium segregation and not the non-equilibrium segregation that develops during cooling or in service. These limits mark where DFT, CALPHAD, magnetic, kinetic or complexion models remain necessary.

\section{Conclusion}
\label{sec:conclusion}

We have developed a macroscopic-atom spectral theory for predicting grain-boundary (GB) segregation in multicomponent alloys without requiring an atomistic calculation for every alloy composition. Elemental properties, including atomic size and surface enthalpy, were mapped onto the local strain and bond deficit of individual GB sites, from which the single-solute segregation energy spectra and the cross-species correlations governing site competition and co-segregation were constructed. Application of the theory to Al-, Fe-, and Mg-based systems shows that the resulting spectral information enables both the ranking of site competition between solute pairs and the prediction of segregation behavior in ternary alloys. The key conclusions are summarized as follows:

\begin{itemize}
	\item Local strain and bond deficit provide a compact and physically interpretable representation of site-wise segregation energetics. Together, these two descriptors account for 91\% of the site-to-site co-variation of 40 solutes in a density functional theory (DFT) Al polycrystal, indicating that both the individual segregation spectra and their cross-species correlations originate from a small number of physically meaningful quantities rather than from solute-specific parameterization.
	
	\item The framework is transferable to hosts excluded from calibration. For such hosts, the predicted spectral means correlate with atomistic results with $r = 0.81$, and the predicted spectra outperform embedded-atom spectra against DFT benchmarks in both Fe and Al. Macroscopic elemental properties can therefore substitute for host-specific atomistic calculations at the screening level, while retaining a clear physical interpretation of the underlying strain and bonding contributions.
	
	\item Site competition between solute species can be ranked without atomistic input for the individual pairs. Across 870 Al solute pairs, the predicted ordering reproduces the atomistic ranking with a Spearman coefficient of 0.90, demonstrating that the theory captures not only the position of each spectrum but also the correlations that determine which species occupy the favorable GB sites.
	
	\item Without any fitting to multicomponent observations, the framework reproduces 11 of 12 clean equilibrium Fe ternaries, 49 of 55 Mg ternaries, and 58 of 81 reported segregation directions outside Mg. This agreement across chemically distinct hosts indicates that the site-competition and co-segregation trends follow directly from the elemental-property-based spectra and their correlations.
	
	\item Predictions recorded before Monte Carlo (MC) simulations of an 82,699-atom polycrystal support the proposed interpretation of multicomponent segregation and delineate its limits. Specific metal--metalloid pair chemistry is identified as the primary source of the observed deviations, indicating the cases in which atomistic resolution remains necessary and where targeted calculations are most efficiently deployed.
\end{itemize}

We conclude that site competition is the baseline behavior of multicomponent GB segregation, that co-segregation requires sufficiently strong chemical attraction between solute species to override this competition, and that bulk compound formation further constrains segregation by limiting the available solute reservoir. Additionally, we have demonstrated the transferability of the macroscopic-atom description across hosts and solute pairs, which allows segregation spectra and their cross-species correlations to be estimated directly from elemental properties. These findings advance the physical understanding of site competition and co-segregation in multicomponent alloys and provide a transferable pathway for screening GB chemistry across complex composition spaces, including steels and high-entropy alloys, and for identifying where targeted atomistic calculations are most valuable.

\section*{CRediT authorship contribution statement}
\textbf{Dennis Boakye}: Writing (original draft), Writing (review and editing), Visualization, Validation, Methodology, Investigation, Formal analysis, Data curation. \textbf{Chuang Deng}: Writing (review and editing), Supervision, Software, Resources, Project administration, Investigation, Funding acquisition, Conceptualization.

\section*{Declaration of competing interest}
The authors declare that they have no known competing financial interests or personal relationships that could have influenced the work reported in this paper.

\section*{Acknowledgments}
This research was supported by the NSERC Alliance International Catalyst (ALLRP 592696-24), Canada, and by the use of a high-performance computing system at the University of Manitoba. During the preparation of this manuscript, the authors used Claude to improve its readability. The manuscript was carefully reviewed and edited as a result of using this tool and take full responsibility for the content of the publication. 

\section*{Data availability}
The data used for this research is available on reasonable request. 

\section*{Supplementary material}
Supplementary material associated with this article contains the list of symbols, the derivation of the isotherm, the cross-validation of model variants, the light-element DFT spectra and correlations in Fe, the MLIP validation and pair energies, the measured anchors and bulk-sink parameters, the complete benchmark with sources, the light-element predictions in other hosts, the full lattice Monte Carlo results, the recorded predictions, and the sensitivity test with computed pair energies.

\scriptsize
\bibliographystyle{elsarticle-num}
\biboptions{sort&compress}
\bibliography{references}

@book{mclean1957grain,
  author = {D. McLean},
  title = {Grain Boundaries in Metals},
  publisher = {Oxford University Press},
  address = {Oxford},
  year = {1957}
}

@book{lejcek2010grain,
  author = {P. Lej\v{c}ek},
  title = {Grain Boundary Segregation in Metals},
  publisher = {Springer},
  address = {Berlin},
  year = {2010},
  doi = {10.1007/978-3-642-12505-8}
}

@article{lejcek2017interfacial,
  author = {P. Lej\v{c}ek and M. \v{S}ob and V. Paidar},
  title = {Interfacial segregation and grain boundary embrittlement: An overview and critical assessment of experimental data and calculated results},
  journal = {Prog. Mater. Sci.},
  volume = {87},
  pages = {83\hyph{}139},
  year = {2017},
  doi = {10.1016/j.pmatsci.2016.11.001},
  publisher = {Elsevier}
}

@article{lejcek2016interstitial,
  author = {P. Lej\v{c}ek and S. Hofmann},
  title = {Interstitial and substitutional solute segregation at individual grain boundaries of $\alpha$-iron: data revisited},
  journal = {J. Phys.: Condens. Matter},
  volume = {28},
  pages = {064001},
  year = {2016},
  doi = {10.1088/0953-8984/28/6/064001},
  publisher = {IOP Publishing}
}

@article{lejvcek2002prediction,
	title={Prediction of enthalpy and entropy of grain boundary segregation},
	author={Lej{\v{c}}ek, P and Hofmann, S},
	journal={Surface and Interface Analysis: An International Journal devoted to the development and application of techniques for the analysis of surfaces, interfaces and thin films},
	volume={33},
	number={3},
	pages={203--210},
	year={2002},
	publisher={Wiley Online Library},
	doi={https://doi.org/10.1002/sia.1203}
}

@article{raabe2014grain,
  author = {D. Raabe and M. Herbig and S. Sandl\"obes and Y. Li and D. Tytko and M. Kuzmina and D. Ponge and P.-P. Choi},
  title = {Grain boundary segregation engineering in metallic alloys: A pathway to the design of interfaces},
  journal = {Curr. Opin. Solid State Mater. Sci.},
  volume = {18},
  pages = {253\hyph{}261},
  year = {2014},
  doi = {10.1016/j.cossms.2014.06.002},
  publisher = {Elsevier}
}

@article{chookajorn2012design,
  author = {T. Chookajorn and H.A. Murdoch and C.A. Schuh},
  title = {Design of stable nanocrystalline alloys},
  journal = {Science},
  volume = {337},
  pages = {951\hyph{}954},
  year = {2012},
  doi = {10.1126/science.1224737},
  publisher = {AAAS}
}

@article{weissmuller1993alloy,
  author = {J. Weissm\"uller},
  title = {Alloy effects in nanostructures},
  journal = {Nanostruct. Mater.},
  volume = {3},
  pages = {261\hyph{}272},
  year = {1993},
  doi = {10.1016/0965-9773(93)90088-S},
  publisher = {Elsevier}
}

@article{kirchheim2002grain,
  author = {R. Kirchheim},
  title = {Grain coarsening inhibited by solute segregation},
  journal = {Acta Mater.},
  volume = {50},
  pages = {413\hyph{}419},
  year = {2002},
  doi = {10.1016/S1359-6454(01)00338-X},
  publisher = {Elsevier}
}

@article{kirchheim1988hydrogen,
  author = {R. Kirchheim},
  title = {Hydrogen solubility and diffusivity in defective and amorphous metals},
  journal = {Prog. Mater. Sci.},
  volume = {32},
  pages = {261\hyph{}325},
  year = {1988},
  doi = {10.1016/0079-6425(88)90010-2},
  publisher = {Elsevier}
}

@book{fowler1939statistical,
  author = {R.H. Fowler and E.A. Guggenheim},
  title = {Statistical Thermodynamics},
  publisher = {Cambridge University Press},
  address = {Cambridge},
  year = {1939}
}

@article{hondros1977theory,
  author = {E.D. Hondros and M.P. Seah},
  title = {The theory of grain boundary segregation in terms of surface adsorption analogues},
  journal = {Metall. Trans. A},
  volume = {8},
  pages = {1363\hyph{}1371},
  year = {1977},
  doi = {10.1007/BF02642850},
  publisher = {Springer}
}

@article{guttmann1975equilibrium,
  author = {M. Guttmann},
  title = {Equilibrium segregation in a ternary solution: A model for temper embrittlement},
  journal = {Surf. Sci.},
  volume = {53},
  pages = {213\hyph{}227},
  year = {1975},
  doi = {10.1016/0039-6028(75)90125-9},
  publisher = {Elsevier}
}

@article{guttmann1995thermochemical,
  author = {M. Guttmann},
  title = {Thermochemical interactions versus site competition in grain boundary segregation and embrittlement in multicomponent systems},
  journal = {J. Phys. IV},
  volume = {5},
  pages = {C7\hyph{}85\hyph{}C7\hyph{}96},
  year = {1995},
  doi = {10.1051/jp4:1995707},
  publisher = {EDP Sciences}
}

@article{wynblatt1977surface,
  author = {P. Wynblatt and R.C. Ku},
  title = {Surface energy and solute strain energy effects in surface segregation},
  journal = {Surf. Sci.},
  volume = {65},
  pages = {511\hyph{}531},
  year = {1977},
  doi = {10.1016/0039-6028(77)90462-9},
  publisher = {Elsevier}
}

@article{miedema1980cohesion,
  author = {A.R. Miedema and P.F. de Ch\^atel and F.R. de Boer},
  title = {Cohesion in alloys - fundamentals of a semi-empirical model},
  journal = {Physica B+C},
  volume = {100},
  pages = {1\hyph{}28},
  year = {1980},
  doi={10.1016/0378-4363(80)90054-6}
}

@book{deboer1988cohesion,
  author = {F.R. de Boer and R. Boom and W.C.M. Mattens and A.R. Miedema and A.K. Niessen},
  title = {Cohesion in Metals: Transition Metal Alloys},
  publisher = {North-Holland},
  address = {Amsterdam},
  year = {1988}
}

@article{white1977spectrum,
  author = {C.L. White and W.A. Coghlan},
  title = {The spectrum of binding energies approach to grain boundary segregation},
  journal = {Metall. Trans. A},
  volume = {8},
  pages = {1403\hyph{}1412},
  year = {1977},
  doi = {10.1007/BF02642853},
  publisher = {Springer}
}

@article{huber2017abinitio,
  author = {L. Huber and B. Grabowski and M. Militzer and J. Neugebauer and J. Rottler},
  title = {Ab initio modelling of solute segregation energies to a general grain boundary},
  journal = {Acta Mater.},
  volume = {132},
  pages = {138\hyph{}148},
  year = {2017},
  doi = {10.1016/j.actamat.2017.04.024},
  publisher = {Elsevier}
}

@article{wagih2019spectrum,
  author = {M. Wagih and C.A. Schuh},
  title = {Spectrum of grain boundary segregation energies in a polycrystal},
  journal = {Acta Mater.},
  volume = {181},
  pages = {228\hyph{}237},
  year = {2019},
  doi = {10.1016/j.actamat.2019.09.034},
  publisher = {Elsevier}
}

@article{wagih2020learning,
  author = {M. Wagih and P.M. Larsen and C.A. Schuh},
  title = {Learning grain boundary segregation energy spectra in polycrystals},
  journal = {Nat. Commun.},
  volume = {11},
  pages = {6376},
  year = {2020},
  doi = {10.1038/s41467-020-20083-6},
  publisher = {Springer Nature}
}

@article{wagih2020grain,
  author = {M. Wagih and C.A. Schuh},
  title = {Grain boundary segregation beyond the dilute limit: Separating the two contributions of site spectrality and solute interactions},
  journal = {Acta Mater.},
  volume = {199},
  pages = {63\hyph{}72},
  year = {2020},
  doi = {10.1016/j.actamat.2020.08.022},
  publisher = {Elsevier}
}

@article{wagih2021thermodynamics,
  author = {M. Wagih and C.A. Schuh},
  title = {Thermodynamics and design of nanocrystalline alloys using grain boundary segregation spectra},
  journal = {Acta Mater.},
  volume = {217},
  pages = {117177},
  year = {2021},
  doi = {10.1016/j.actamat.2021.117177},
  publisher = {Elsevier}
}

@article{wagih2022learning,
  author = {M. Wagih and C.A. Schuh},
  title = {Learning grain-boundary segregation: From first principles to polycrystals},
  journal = {Phys. Rev. Lett.},
  volume = {129},
  pages = {046102},
  year = {2022},
  doi = {10.1103/PhysRevLett.129.046102},
  publisher = {American Physical Society}
}

@article{wagih2023spectrum,
  author = {M. Wagih and C.A. Schuh},
  title = {The spectrum of interstitial solute energies in polycrystals},
  journal = {Scr. Mater.},
  volume = {235},
  pages = {115631},
  year = {2023},
  doi = {10.1016/j.scriptamat.2023.115631},
  publisher = {Elsevier}
}

@article{wagih2025designing,
  author = {M. Wagih and Y. Naunheim and T. Lei and C.A. Schuh},
  title = {Designing for cooperative grain boundary segregation in multicomponent alloys},
  journal = {Proc. Natl. Acad. Sci. U.S.A.},
  volume = {122},
  pages = {e2511930122},
  year = {2025},
  doi = {10.1073/pnas.2511930122},
  publisher = {National Academy of Sciences}
}

@article{matson2026overview,
  author = {T.P. Matson and N. Tuchinda and C.A. Schuh},
  title = {Overview: The spectral model of grain boundary segregation},
  journal = {Acta Mater.},
  volume = {313},
  pages = {122109},
  year = {2026},
  doi = {10.1016/j.actamat.2026.122109},
  publisher = {Elsevier}
}

@article{matson2021atomistic,
  author = {T.P. Matson and C.A. Schuh},
  title = {Atomistic assessment of solute-solute interactions during grain boundary segregation},
  journal = {Nanomaterials},
  volume = {11},
  pages = {2360},
  year = {2021},
  doi = {10.3390/nano11092360},
  publisher = {MDPI}
}

@misc{matson2024grain,
  author = {T.P. Matson and C.A. Schuh},
  title = {Grain boundary segregation and solute-solute interaction energies of binary alloys from atomistic simulations},
  howpublished = {Mendeley Data, V1, doi:10.17632/hrv39bz4k6.1},
  year = {2024},
  doi = {10.17632/hrv39bz4k6.1}
}

@article{marchiy2025spectral,
  author = {G. Marchiy and D. Samsonov and E. Mukhin},
  title = {Spectral model for grain boundary segregation in systems with strong solute-solute interactions},
  journal = {Acta Mater.},
  volume = {294},
  pages = {121044},
  year = {2025},
  doi = {10.1016/j.actamat.2025.121044},
  publisher = {Elsevier}
}

@misc{marchiy2026multicomponent,
  author = {G. Marchiy and F. Kuznetsov and D. Samsonov and E. Mukhin},
  title = {Multicomponent grain boundary segregation dilute-limit model and its effect on nanocrystalline stability},
  howpublished = {arXiv:2606.24534},
  year = {2026},
  doi = {https://doi.org/10.48550/arXiv.2606.24534}
}

@article{zhang2026predicting,
  author = {Z. Zhang and C. Deng},
  title = {Predicting co-segregation in multicomponent alloys with solute-solute interactions},
  journal = {Acta Mater.},
  volume = {318},
  pages = {122628},
  year = {2026},
  doi = {10.1016/j.actamat.2026.122628},
  publisher = {Elsevier}
}

@article{boakye2027rapid,
  title = {Rapid modeling of segregation-driven metal–oxide adhesion in high-entropy alloys using macroscopic atom model},
  author = {D. Boakye and C. Deng},
  journal = {Acta Materialia},
  volume = {322},
  pages = {122840},
  year = {2027},
  issn = {1359-6454},
  url = {https://www.sciencedirect.com/science/article/pii/S1359645426009389},
  doi = {10.1016/j.actamat.2026.122840},
  publisher = {Elsevier}
}

@article{wu2025unconventional,
  author = {H. Wu and X. Li and W. Gao and Q. Jiang},
  title = {Unconventional solute-matrix bonding at metallic polycrystal grain boundaries and their amorphous counterparts},
  journal = {Acta Mater.},
  volume = {301},
  pages = {121590},
  year = {2025},
  doi = {10.1016/j.actamat.2025.121590},
  publisher = {Elsevier}
}

@article{li2024analytic,
  author = {X. Li and Y. Li and W. Gao},
  title = {An analytic descriptor for determining the effect of grain-boundary structures of metals on solute segregation},
  journal = {J. Appl. Phys.},
  volume = {135},
  pages = {145303},
  year = {2024},
  doi={https://doi.org/10.1063/5.0198079}
}

@article{li2025universal,
  author = {X. Li and P.T. Liu and W. Gao and X.Q. Chen and Q. Jiang},
  title = {A universal descriptor to determine the effect of solutes in segregation at grain boundaries},
  journal = {J. Mater. Sci. Technol.},
  volume = {222},
  pages = {22\hyph{}27},
  year = {2025},
  doi = {10.1016/j.jmst.2024.10.014},
  publisher = {Elsevier}
}

@article{li2026alloying,
  author = {X. Li and J. Cheng and W. Gao},
  title = {Alloying effects of co-segregation at metal grain boundaries: from analytic model to transferable machine-learning framework},
  journal = {Comput. Mater. Sci.},
  volume = {272},
  pages = {114884},
  year = {2026},
  doi={https://doi.org/10.1016/j.commatsci.2026.114884}
}

@article{zhu2024model,
  author = {Q. Zhu and X. Li and W. Gao and Q. Jiang},
  title = {A model for predicting grain-boundary energy of refractory high-entropy alloys based on local element concentration},
  journal = {Scr. Mater.},
  volume = {252},
  pages = {116285},
  year = {2024},
  doi = {10.1016/j.scriptamat.2024.116285},
  publisher = {Elsevier}
}

@article{li2020grain,
  author = {L. Li and R.D. Kamachali and Z. Li and Z. Zhang},
  title = {Grain boundary energy effect on grain boundary segregation in an equiatomic high-entropy alloy},
  journal = {Phys. Rev. Mater.},
  volume = {4},
  pages = {053603},
  year = {2020},
  doi = {10.1103/PhysRevMaterials.4.053603},
  publisher = {American Physical Society}
}

@article{mccarthy2021emergence,
  author = {M.J. McCarthy and T.J. Rupert},
  title = {Emergence of near-boundary segregation zones in face-centered cubic multiprincipal element alloys},
  journal = {Phys. Rev. Mater.},
  volume = {5},
  pages = {113601},
  year = {2021},
  doi = {10.1103/PhysRevMaterials.5.113601},
  publisher = {American Physical Society}
}

@article{xing2018solute,
  author = {W. Xing and A.R. Kalidindi and D. Amram and C.A. Schuh},
  title = {Solute interaction effects on grain boundary segregation in ternary alloys},
  journal = {Acta Mater.},
  volume = {161},
  pages = {285\hyph{}294},
  year = {2018},
  doi = {10.1016/j.actamat.2018.09.005},
  publisher = {Elsevier}
}

@article{ito2022first,
  author = {K. Ito and H. Sawada},
  title = {First-principles analysis of the grain boundary segregation of transition metal alloying elements in $\gamma$Fe},
  journal = {Comput. Mater. Sci.},
  volume = {210},
  pages = {111050},
  year = {2022},
  doi = {10.1016/j.commatsci.2021.111050},
  publisher = {Elsevier}
}

@article{darling2011stabilized,
  author = {K.A. Darling and B.K. VanLeeuwen and J.E. Semones and C.C. Koch and R.O. Scattergood and L.J. Kecskes and S.N. Mathaudhu},
  title = {Stabilized nanocrystalline iron-based alloys: Guiding efforts in alloy selection},
  journal = {Mater. Sci. Eng. A},
  volume = {528},
  pages = {4365\hyph{}4371},
  year = {2011},
  doi = {10.1016/j.msea.2011.02.080},
  publisher = {Elsevier}
}

@article{mai2022segregation,
  author = {H.L. Mai and X.Y. Cui and D. Scheiber and L. Romaner and S.P. Ringer},
  title = {The segregation of transition metals to iron grain boundaries and their effects on cohesion},
  journal = {Acta Mater.},
  volume = {231},
  pages = {117902},
  year = {2022},
  doi = {10.1016/j.actamat.2022.117902},
  publisher = {Elsevier}
}

@misc{mai2026grain,
  author = {H.L. Mai and X.Y. Cui and T. Hickel and S.P. Ringer and J. Neugebauer},
  title = {Grain boundary segregation of light elements and their effects on cohesion in ferritic steels},
  howpublished = {arXiv:2605.01181},
  year = {2026},
  doi = {10.48550/arXiv.2605.01181}
}

@article{erhart1981equilibrium,
  author = {H. Erhart and H.J. Grabke},
  title = {Equilibrium segregation of phosphorus at grain boundaries of {Fe-P}, {Fe-C-P}, {Fe-Cr-P}, and {Fe-Cr-C-P} alloys},
  journal = {Met. Sci.},
  volume = {15},
  pages = {401\hyph{}408},
  year = {1981},
  doi = {10.1179/030634581790426877},
  publisher = {Maney Publishing}
}

@article{grabke1987effects,
  author = {H.J. Grabke and R. M\"oller and H. Erhart and S.S. Brenner},
  title = {Effects of the alloying elements {Ti}, {Nb}, {Mo} and {V} on the grain boundary segregation of {P} in iron and steels},
  journal = {Surf. Interface Anal.},
  volume = {10},
  pages = {202\hyph{}209},
  year = {1987},
  doi = {10.1002/sia.740100405},
  publisher = {Wiley}
}

@article{briant1980effects,
  author = {C.L. Briant},
  title = {The Effects of Sulfur and Phosphorus on the Intergranular Corrosion of 304 Stainless Steel},
  journal = {Acta Metall.},
  volume = {36},
  pages = {497\hyph{}509},
  year = {1988},
  doi={https://doi.org/10.5006/0010-9312-36.9.497}
}

@article{tauber1978grain,
  author = {G. Tauber and H.J. Grabke},
  title = {Grain boundary segregation of sulfur, nitrogen, and carbon in $\alpha$-iron},
  journal = {Ber. Bunsenges. Phys. Chem.},
  volume = {82},
  pages = {298\hyph{}302},
  year = {1978},
  doi = {10.1002/bbpc.197800042},
  publisher = {Wiley}
}

@article{oikawa2000thermodynamic,
  author = {K. Oikawa and H. Mitsui and H. Ohtani and K. Ishida},
  title = {Thermodynamic calculations of phase equilibria in the {Fe-Cr-S} system},
  journal = {ISIJ Int.},
  volume = {40},
  pages = {182\hyph{}190},
  year = {2000},
  doi = {10.2355/isijinternational.40.182},
  publisher = {Iron and Steel Institute of Japan}
}

@article{tokunaga2009thermodynamic,
  author = {T. Tokunaga and N. Hanaya and H. Ohtani and M. Hasebe},
  title = {Thermodynamic analysis of phase equilibria in the {Fe-Nb-P} ternary system},
  journal = {ISIJ Int.},
  volume = {49},
  pages = {947\hyph{}956},
  year = {2009},
  doi = {10.2355/isijinternational.49.947},
  publisher = {Iron and Steel Institute of Japan}
}

@incollection{friedel1969transition,
  author = {J. Friedel},
  title = {Transition metals. Electronic structure of the d-band. Its role in the crystalline and magnetic structures},
  booktitle = {The Physics of Metals},
  editor = {J.M. Ziman},
  publisher = {Cambridge University Press},
  address = {Cambridge},
  pages = {340\hyph{}408},
  year = {1969}
}

@inproceedings{batatia2022mace,
  author = {I. Batatia and D.P. Kov\'acs and G.N.C. Simm and C. Ortner and G. Cs\'anyi},
  title = {{MACE}: Higher order equivariant message passing neural networks for fast and accurate force fields},
  booktitle = {Advances in Neural Information Processing Systems},
  volume = {35},
  pages = {11423\hyph{}11436},
  year = {2022},
  doi={10.52202/068431-0830}
}

@article{batatia2025foundation,
	title={A foundation model for atomistic materials chemistry},
	author={Batatia, Ilyes and Benner, Philipp and Chiang, Yuan and Elena, Alin M and Kov{\'a}cs, D{\'a}vid P and Riebesell, Janosh and Advincula, Xavier R and Asta, Mark and Avaylon, Matthew and Baldwin, William J and others},
	year={2025},
	publisher={Newcastle University},
	doi={10.1063/5.0297006}
}

@misc{kaplan2025foundational,
  author = {A.D. Kaplan and R. Liu and J. Qi and T.W. Ko and B. Deng and J. Riebesell and G. Ceder and K.A. Persson and S.P. Ong},
  title = {A foundational potential energy surface dataset for materials},
  howpublished = {arXiv:2503.04070},
  year = {2025},
  doi = {10.48550/arXiv.2503.04070}
}

@article{fan2021neuroevolution,
  author = {Z. Fan and Z. Zeng and C. Zhang and Y. Wang and K. Song and H. Dong and Y. Chen and T. Ala-Nissila},
  title = {Neuroevolution machine learning potentials: Combining high accuracy and low cost in atomistic simulations and application to heat transport},
  journal = {Phys. Rev. B},
  volume = {104},
  pages = {104309},
  year = {2021},
  doi = {10.1103/PhysRevB.104.104309},
  publisher = {American Physical Society}
}

@article{xu2025gpumd,
  author = {K. Xu and others},
  title = {{GPUMD} 4.0: A high-performance molecular dynamics package for versatile materials simulations with machine-learned potentials},
  journal = {Mater. Genome Eng. Adv.},
  volume = {3},
  pages = {e70028},
  year = {2025},
  doi = {10.1002/mgea.70028},
  publisher = {Wiley}
}

@article{larsen2016robust,
  author = {P.M. Larsen and S. Schmidt and J. Schi{\o}tz},
  title = {Robust structural identification via polyhedral template matching},
  journal = {Model. Simul. Mater. Sci. Eng.},
  volume = {24},
  pages = {055007},
  year = {2016},
  doi = {10.1088/0965-0393/24/5/055007},
  publisher = {IOP Publishing}
}

@article{mendelev2003development,
  author = {M.I. Mendelev and S. Han and D.J. Srolovitz and G.J. Ackland and D.Y. Sun and M. Asta},
  title = {Development of new interatomic potentials appropriate for crystalline and liquid iron},
  journal = {Philos. Mag.},
  volume = {83},
  pages = {3977\hyph{}3994},
  year = {2003},
  doi = {10.1080/14786430310001613264},
  publisher = {Taylor \& Francis}
}

@article{hu2024computational,
  author = {C. Hu and R. Dingreville and B.L. Boyce},
  title = {Computational modeling of grain boundary segregation: A review},
  journal = {Comput. Mater. Sci.},
  volume = {232},
  pages = {112596},
  year = {2024},
  doi = {10.1016/j.commatsci.2023.112596},
  publisher = {Elsevier}
}

@article{murdoch2013estimation,
  author = {H.A. Murdoch and C.A. Schuh},
  title = {Estimation of grain boundary segregation enthalpy and its role in stable nanocrystalline alloy design},
  journal = {Journal of Materials Research},
  volume = {28},
  number = {16},
  pages = {2154\hyph{}2163},
  year = {2013},
  doi = {10.1557/jmr.2013.211},
  publisher = {Cambridge University Press}
}

@article{zhang2016miedema,
  author = {R.F. Zhang and S.H. Zhang and Z.J. He and J. Jing and S.H. Sheng},
  title = {Miedema Calculator: A thermodynamic platform for predicting formation enthalpies of alloys within framework of Miedema's Theory},
  journal = {Computer Physics Communications},
  volume = {209},
  pages = {58\hyph{}69},
  year = {2016},
  doi = {10.1016/j.cpc.2016.08.013},
  publisher = {Elsevier}
}

\end{document}